\documentclass[10pt]{elsarticle}
 
\usepackage{calrsfs,graphicx,array,natbib,enumitem}
\usepackage{setspace,wrapfig,marvosym,bbm,stmaryrd,centernot,color,url,amssymb}
\usepackage{amsmath}

\newtheorem{rem}{Remark}[section]
\newtheorem{theorem}{Theorem}[section]
\newtheorem{assumption}{Assumption}[section]

\newlength{\kaka}

\newcommand{\ahref}[2]{}

\newcommand{\obs}{^{\text{obs}}}

\newcommand{\R}{\mathbb{R}}

\newcommand{\ff}{^{\text{\tiny f}}}

\newcommand{\beq}{\begin{equation}}
\newcommand{\eeq}{\end{equation}}
\newcommand{\lb}{\label}

\newcommand{\bea}{\begin{eqnarray}}
\newcommand{\eea}{\end{eqnarray}}
\newcommand{\bxr}{\begin{array}}
\newcommand{\exr}{\end{array}}

\newcommand\exs{\hspace*{0.4mm}}

\newcommand\nxs{\hspace*{-0.2mm}}

\newcommand{\norms}[1]{\parallel\! #1 \!\parallel}

\newcommand{\bC} {\boldsymbol{C}}
\newcommand{\bK} {\boldsymbol{K}}
\newcommand{\bV} {\boldsymbol{V}}
\newcommand{\bR} {\boldsymbol{\sf R}}

\newcommand{\bn} {\boldsymbol{n}}
\newcommand{\ba} {\boldsymbol{a}}

\newcommand{\bx} {\boldsymbol{x}}
\newcommand{\by} {\boldsymbol{y}}

\newcommand{\bg} {{\boldsymbol{g}}}

\newcommand{\bI} {\boldsymbol{I}}

\newcommand{\sip} {\!\cdot\!}

\newcommand{\bzero}{\boldsymbol{0}}

\newcommand{\bu} {\boldsymbol{u}}

\newcommand{\bt} {{\boldsymbol{t}}}

\newcommand{\bv} {\boldsymbol{v}}

\newcommand{\bxi} {\boldsymbol{\xi}}

\newcommand{\bw} {\boldsymbol{w}}

\newcommand{\bPhi}{\boldsymbol{\Phi}}

\newcommand{\dualGA}[2]{\left< #1, #2 \right>_{\Gamma}}

\begin{document}

\begin{frontmatter}

\title{Laboratory application of sampling approaches to inverse scattering}

\author{\corref{cor1}{Fatemeh Pourahmadian$^{1,2}$}}  \author{Hao Yue$^{1}$}

\address{$^{1}$Department of Civil, Environmental \& Architectural Engineering, University of Colorado Boulder}
\address{$^2$ Department of Applied Mathematics, University of Colorado Boulder, USA}

\cortext[cor1]{Corresponding author: tel. 303-492-2027, email {\tt fatemeh.pourahmadian@colorado.edu}}

\date{\today}

\begin{abstract}

This study presents an experimental investigation of the recently established generalized linear sampling method (GLSM)~\cite{Fatemeh2017} for non-destructive evaluation of damage in elastic materials. To this end, ultrasonic shear waves are generated in a prismatic slab of charcoal granite featuring a discontinuity interface induced by the three-point bending (3PB). The interaction of probing waves with the 3PB-induced damage gives rise to transient velocity responses measured on the sample's boundary by a 3D scanning laser Doppler vibrometer. Thus obtained waveform data are then carefully processed to retrieve the associated spectra of scattered displacement fields. On deploying multifrequency sensory data, the GLSM indicators are computed and their counterparts associated with the classical linear sampling method (LSM)~\cite{Fiora2003} for comparative analysis. Verified with in-situ observations, the GLSM map successfully exposes the support of hidden scatterers in the specimen with a remarkable clarity and resolution compared to its predecessor LSM. It is further shown that the GLSM remains robust for sparse and partial-aperture data inversion, thanks to its rigorous formulation. For completeness, the one-sided reconstruction by both indicators is investigated. 

\end{abstract}


\end{frontmatter}

\section{Introduction} \label{sec1}

Inverse scattering solutions are sought for uncovering geometrical and physical properties of hidden objects in a medium from remote (or boundary) observations of thereby scattered waveforms. In this context, waveform tomography of discontinuity surfaces bear direct relevance to (a) timely detection of degradation in safety-sensitive components, (b) in-situ monitoring of additive manufacturing processes, and (c) efficient energy mining from unconventional resources. Existing optimization-based approaches to waveform inversion typically incur high computational cost as a crucial obstacle to real-time sensing. Lately, non-iterative inverse scattering solutions~\cite{cako2016,bonn2019,Fatemeh2017} have been brought under the spotlight for their capabilities pertinent to fast imaging in highly scattering media~\cite{pour2019}. Spurred by the early study in~\cite{Kress1995}, such developments include: (i) the Factorization Method (FM)~\cite{Kirsch2008, Bouk2013}, (ii) the Linear Sampling Method (LSM)~\cite{Fiora2003,cako2016}, (iii) MUSIC algorithms~\cite{Park2015(2)}, (iv) the method of Topological Sensitivity (TS)~\cite{Fatemeh2015,Fatemeh2015(2)}, and (v) the Generalized Linear Sampling Method (GLSM)~\cite{audi2017,Fatemeh2017}. Among these, the FM, LSM, and GLSM  inherently carry a superior localization property that potentially leads to high-fidelity geometric reconstruction. 

This study is focused on the GLSM indicator~\cite{Fatemeh2017,audi2017} developed by building upon the factorization method and recent theories on design of imaging functionals~\cite{audi2017, Fatemeh2017(2)}. More specifically, the GLSM is a non-iterative, full-waveform approach to elastic-wave imaging of 3D discontinuity surfaces with non-trivial (generally heterogeneous and dissipative) interfacial condition. This indicator map -- targeting geometric reconstruction of extended interfaces -- is shown to be (a) agnostic with respect to the contact condition at the interface, (b) robust against measurement errors, and (c) flexible in terms of sensing parameters, e.g. the illumination frequency. 

On the verification side, the effectiveness of sampling methods for elastic waveform tomography has been extensively examined by numerical simulations, see e.g.,~\cite{cako2016,audi2017,Fatemeh2017,pour2019}. A systematic experimental investigation of these imaging tools, however, is still lacking. To help bridge the gap, a few recent studies~\cite{baro2016,baro2018} demonstrate successful performance of the classical linear sampling method in a laboratory setting. The present work augments these efforts by investigating the generalized linear sampling technique in an experimental campaign for the shape reconstruction of an extended damage zone from boundary data. In primary experiments, ultrasonic waves are induced in an \emph{intact} slab of charcoal granite and the resulting velocity responses are captured by a 3D scanning laser Doppler vibrometer over the sample's edges, furnishing the incident fields affiliated with every source location. The sample is then notched and fractured in the three-point-bending (3PB) configuration, then probed by ultrasonic waves in a similar fashion as in the primary experiments. The secondary measurements carry the scattering signature of 3PB-induced damage in the granite.     
The primary and secondary sensory data are then carefully processed and transformed into the frequency domain to compute the GLSM and LSM indicators and recover the support of damage zone. In this study, the data inversion is adapted to the testing configuration and the nature of measured waveforms. In particular, the reconstruction procedure is reformulated for multi-frequency inverse scattering, also the GLSM cost functional is carefully modified to accommodate for a highly asymmetric scattering operator  resulting from the sparse sampling (of the incident surface). It is shown that the GLSM indicator successfully reconstructs the process zone's geometry including the pre-manufactured notch and the (heterogeneous) mode I fracture induced by three-point bending. The performance of LSM and GLSM imaging functionals are compared. The influences of key testing parameters on the fidelity of reconstruction -- including the source/measurement aperture and sensing resolution are also investigated for both indicators.

This paper is organized as follows. \textcolor{black}{Section}~\ref{prelim} formulates the direct scattering problem within the context of laboratory experiments, and provides an overview of the data inversion platform. \textcolor{black}{Section}~\ref{exp_set} describes the experimental procedure and showcases the ``raw" measurements. \textcolor{black}{Section}~\ref{DSP} includes a detailed account of signal processing in time and space in preparation for data inversion. \textcolor{black}{Section}~\ref{DI} computes the (generalized) linear sampling functionals using multi-frequency data. \textcolor{black}{Section}~\ref{RE} presents and discusses the results.

\section{Theoretical foundation} \label{prelim}
   
This section briefly outlines two theories of inverse scattering considered in this study -- namely, the classical linear sampling method~\cite{cako2016,Has2013} and the recently developed generalized linear sampling technique~\cite{Fatemeh2017,Audibert2014}.        

\vspace{-1mm}
\subsection{Problem statement} \lb{FP}

Let $\mathcal{B} \subset \R^d$, $d = 2, 3$, denote a finite elastic body characterized by mass density $\rho$, and Lam\'{e} parameters $\mu$ and~$\lambda$, which henceforth is referred to as the \emph{baseline model}. A set of unknown discontinuities $\Gamma$ is embedded in $\mathcal{B}$ whose support is possibly disjoint and of arbitrary shape. More specifically, $\Gamma$ may be decomposed into $N$ smooth open subsets $\Gamma_n$, each of which may be arbitrarily extended to a closed Lipschitz surface $\partial \text{\sf D}_n$ enclosing a bounded simply connected domain $\text{\sf D}_n \subset \R^d$, so that $\Gamma \!=\! {\textstyle \bigcup_{n =1}^{N}} \Gamma_n \subset {\textstyle \bigcup_{n =1}^{N}}  \partial \text{\sf D}_n$. The contact at the surface of $\Gamma$ is characterized by a symmetric and heterogeneous interfacial stiffness matrix $\bK(\bxi), \, \bxi \in \Gamma$, synthesizing the spatially varying nature of rough interfaces. Here, $\bK$ is arbitrary and a priori unknown. 
\vspace{-1 mm}
\begin{assumption}\label{nodiss}
In this study, the interfacial energy dissipation on $\Gamma$ is assumed negligible during the course of ultrasonic measurements. This may be justified owing to the small amplitude of motion, and short period of observation in the experimental campaign.    
\end{assumption}
\vspace{-1 mm}
The domain $\mathcal{B}$ is excited by an ultrasonic source on its external boundary $\partial \mathcal{B}$ so that the corresponding incident field $\bu^{\textrm{f}}(\bxi,t)$ in the \emph{baseline model} is governed by 
\vspace{-1.5 mm}
\beq\lb{uf}
\begin{aligned}
&\nabla \exs\sip\exs [\bC \exs \colon \! \nabla \bu^{\textrm{f}}\exs](\bxi,t) \,-\, \rho \exs \ddot{\bu}^{\textrm{f}}(\bxi,t) ~=~ \bzero, \quad &  \big(\bxi \in {\mathcal{B}}, t \in (0,T] \big) \\*[0.5mm]
&\bn \exs\sip\exs \bC \exs \colon \!  \nabla  \bu^{\textrm{f}}(\bxi,t)~=~\bg(\bxi,t),  \quad & \big(\bxi \in \partial {\mathcal{B}}_t, t \in (0,T] \big) \\*[0.5mm] 
& \bu^{\textrm{f}}(\bxi,t)~=~\bzero,   \quad & \big(\bxi \in \partial\mathcal{B}_u, t \in (0,T] \big)\\*[0.5mm] 
& \bu^{\textrm{f}}(\bxi,0)~=~ \dot{\bu}^{\textrm{f}}(\bxi,0)~=~ \bzero,   \quad & \big(\bxi \in \overline{\mathcal{B}}, t = 0 \big)
\end{aligned}   
\vspace{-1.5 mm}
\eeq  
where the fourth-order elasticity tensor $\bC = \lambda\bI_2\!\otimes\!\bI_2 + 2\mu\bI_4$ with $\bI_m \,(m\!=\!2,4)$ denoting the $m$th-order symmetric identity tensor; the single and double over-dots indicate first- and second- order time derivates, respectively; $T$ signifies the testing interval; $\bn$ is the unit outward normal to the sample's boundary $\partial \mathcal{B}$; $\bg(\bxi,t)$ represents the external traction on the Neumann part of the boundary $\partial {\mathcal{B}}_t \subset \partial {\mathcal{B}}$ which includes the source input; the displacement vanishes on the boundary's Dirichlet part $\partial\mathcal{B}_u \subset \partial\mathcal{B}$; and, overline indicates the closure of a set e.g.,~$\overline{\mathcal{B}} = {\mathcal{B}} \cup \partial {\mathcal{B}}$. The interaction of $\bu^{\textrm{f}}$ with the hidden scatterers $\Gamma$ gives rise to the \emph{total field} $\bu(\bxi,t)$ in the physical domain satisfying      
\vspace{-1.5 mm}
\beq\lb{uk}
\begin{aligned}
&\nabla \exs\sip\exs [\bC \exs \colon \! \nabla \bu\exs](\bxi,t) \,-\, \rho \exs \ddot{\bu}(\bxi,t) ~=~ \bzero, \quad &  \big(\bxi \in {\mathcal{B}}\backslash \Gamma, t \in (0,T] \big) \\*[0.5mm]
&\bn_{{\small \Gamma}} \sip\exs \bC \exs \colon \!  \nabla  \bu(\bxi,t)~=~\bK(\bxi) \llbracket \bu \rrbracket(\bxi,t),  \quad & \big(\bxi \in \Gamma, t \in (0,T] \big) \\*[0.5mm]
&\bn \exs\sip\exs \bC \exs \colon \!  \nabla  \bu(\bxi,t)~=~\bg(\bxi,t),  \quad & \big(\bxi \in \partial {\mathcal{B}}_t, t \in (0,T] \big) \\*[0.5mm] 
& \bu(\bxi,t)~=~\bzero,   \quad & \big(\bxi \in \partial\mathcal{B}_u, t \in (0,T] \big)\\*[0.5mm] 
& \bu(\bxi,0)~=~ \dot{\bu}(\bxi,0)~=~ \bzero,   \quad & \big(\bxi \in \overline{\mathcal{B}}, t = 0 \big)
\end{aligned}   
\vspace{-0.5 mm}
\eeq  
where $\llbracket \bu \rrbracket(\bxi,t)$ indicates the jump in displacement field across $\bxi \in \Gamma$; $\bn_{{\small \Gamma}}$ indicates the unit normal vector on $\Gamma$ which on recalling $\Gamma \subset {\textstyle \bigcup_{n =1}^{N}}  \partial \text{\sf D}_n$, is outward to $\text{\sf D}_n$. The wave motion is measured in terms of $\bu(\bxi,t)$ over the observation surface $\bxi \in S\obs\subset \partial {\mathcal{B}}_t$, and the corresponding scattered field may be computed as
\vspace{-1.5 mm}
\beq\lb{SDF}
\bv(\bxi,t) := [\bu- \bu^{\textrm{f}}\exs](\bxi,t),  
\vspace{-2 mm}
\eeq  
satisfying
\beq\lb{bv}
\begin{aligned}
&\nabla \exs\sip\exs [\bC \exs \colon \! \nabla \bv\exs](\bxi,t) \,-\, \rho \exs \ddot{\bv}(\bxi,t) ~=~ \bzero, \quad &  \big(\bxi \in {\mathcal{B}}\backslash \Gamma, t \in (0,T] \big) \\*[0.5mm]
&\bn_{{\small \Gamma}} \sip\exs \bC \exs \colon \!  \nabla  \bv(\bxi,t)~=~\bK(\bxi) \llbracket \bv \rrbracket(\bxi,t) - \bt^{\textrm{f}}(\bxi,\omega),  \quad & \big(\bxi \in \Gamma, t \in (0,T] \big) \\*[0.5mm]
&\bn \exs\sip\exs \bC \exs \colon \!  \nabla  \bv(\bxi,t)~=~\bzero,  \quad & \big(\bxi \in \partial {\mathcal{B}}_t, t \in (0,T] \big) \\*[0.5mm]
& \bv(\bxi,t)~=~\bzero,   \quad & \big(\bxi \in \partial\mathcal{B}_u, t \in (0,T] \big)\\*[0.5mm]
& \bv(\bxi,0)~=~ \dot{\bv}(\bxi,0)~=~ \bzero,   \quad & \big(\bxi \in \overline{\mathcal{B}}, t = 0 \big)
\end{aligned}   
\vspace{-0.5 mm}
\eeq 
where $\bt^{\textrm{f}} = \bn_{{\small \Gamma}} \exs\sip\exs \bC \exs \colon \!  \nabla  \bu^{\textrm{f}}$ is the free-field traction on the surface of $\Gamma$. The experiments are repeated for a set of ultrasonic excitations on the incident surface $S^{\textrm{inc}\!}\! \subset \partial {\mathcal{B}}_t$.

To assist the inverse analysis, let us introduce the relevant function spaces as the following,
\vspace*{-1.5mm}
\beq\lb{funS2}
\begin{aligned}
&H^{\pm \frac{1}{2}}(\Gamma) ~:=~\big\lbrace f\big|_{\Gamma} \! \colon \,\,\, f \in H^{\pm \frac{1}{2}}(\partial \text{\sf D}) \big\rbrace, \\*[0.0 mm]
& \tilde{H}^{\pm \frac{1}{2}}(\Gamma) ~:=~\big\lbrace  f \in H^{\pm\frac{1}{2}}(\partial \text{\sf D}) \colon  \,\,\, \text{supp}(f) \subset \overline{\Gamma} \exs \big\rbrace,
\end{aligned}
\vspace*{-1.5mm}
\eeq
where $\text{\sf D} = {\textstyle \bigcup_{n =1}^{N}} \text{\sf D}_n$ is a multiply connected Lipschitz domain of bounded support such that $\Gamma \subset \partial \text{\sf D}$, and $\overline{\Gamma} \colon \!\!\! =  \Gamma \cup \partial\Gamma$ denotes the closure of $\Gamma \!=\! {\textstyle \bigcup_{n =1}^{N}} \Gamma_n$. Recall that every $\Gamma_n$ is an open set (relative to $\partial \text{\sf D}_n$) with a positive surface measure. Note that since $\bv \in  H^1({\mathcal{B}}\backslash\Gamma)^3$, then by trace theorems  $\llbracket \bv \rrbracket\in\tilde{H}^{1/2}(\Gamma)^3$.

\vspace{-1mm}
\subsection{Inverse solution} \lb{InvS}

The (generalized) linear sampling indicators use the spectrum of scattered displacement field $\bv$ on $S\obs$ to \emph{non-iteratively} reconstruct the support of hidden scatterers $\Gamma$ via synthetic wavefront shaping. To this end, the scattering operator $\Lambda:\, L^2(S^{\text{inc}})^3\nxs\times L^2(\Omega)^3  \exs\to\exs L^2(S\obs)^3\nxs\times L^2(\Omega)^3$ is constructed over a frequency bandwidth $\Omega := [\omega_{\min}\,\,\, \omega_{\max}] \subset \R^+$ from test data as the following
\vspace{-1.5mm}
\beq\lb{So} 
\Lambda(\bg)(\bxi,\omega) ~=\,  \int_{S^{\text{inc}\!}} \bV(\bxi,\by;\omega) \sip \bg(\by,\omega) \,\, \text{d}S_{\by}, \qquad \bg \in L^2(S^{\text{inc}})^3\nxs\times L^2(\Omega)^3, \quad\!\! \bxi \in S\obs, \,\,\, \omega \in \Omega.
\vspace{-1.5mm}
\eeq 
On denoting by $F(\cdot)$ the Fourier transform operator, $V_{ij}(\bxi,\by;\omega)$, $i,j\!=\!1,2,3$, in~\eqref{So} indicates the $i^{\textrm{th}}$ component of the Fourier transformed displacement $F(\bv)(\bxi,\omega) \in L^2(S\obs)^3\nxs\times L^2(\Omega)^3$ measured at $\bxi \in S\obs$ with frequency $\omega \in \Omega$ due to excitation at $\by \in S^{\textrm{inc}}$ in the $j^{\textrm{th}}$ direction. 

In addition, let us consider the search volume $\mathcal{S} \subset \mathcal{B} \subset \R^{d}$ in the (intact) \emph{baseline model}, and define a set of trial dislocations $L(\bx_\circ, \bR) \subset \mathcal{S}$ such that for every pair $(\bx_\circ,\bR)$, $L\colon\!\!=\bx_\circ\!+\bR{\sf L}$ specifies a smooth arbitrary-shaped fracture~$\sf L$~at $\bx_\circ \subset \mathcal{S}$ whose orientation is identified by a unitary rotation matrix $\bR\!\in\!U(3)$. In this setting, the scattering pattern $\bPhi_L \colon \tilde{H}^{1/2}(L)^3 \nxs\times L^2(\Omega)^3 \rightarrow L^2(S\obs)^3 \nxs\times L^2(\Omega)^3$ on $S\obs$ -- generated by $L(\bx_\circ, \bR)$, as a sole scatterer in $\mathcal{B}$, endowed with an admissible displacement density $\ba(\bxi,\omega)\!\in\!\tilde{H}^{1/2}(L)^3 \nxs\times L^2(\Omega)^3$ -- is governed by 
\vspace{-1.5mm}
\beq\lb{PhiL}
\begin{aligned}
&\nabla \nxs\cdot [\bC \exs \colon \! \nabla \bPhi_L](\bxi,\omega) \,+\, \rho \exs \omega^2\bPhi_L(\bxi,\omega)~=~\bzero, \quad & \big(\bxi \in {\mathcal{B}}\backslash L, \omega \in \Omega \big) \\*[0.5mm]
&\bn \nxs\cdot \bC \exs \colon \!  \nabla  \bPhi_L(\bxi,\omega)~=~\bzero,  \quad & \big(\bxi \in \partial{\mathcal{B}}_t, \omega \in \Omega \big) \\*[0.5mm]
&\bPhi_L(\bxi,\omega)~=~\bzero,  \quad & \big(\bxi \in \partial{\mathcal{B}}_u, \omega \in \Omega \big) \\*[0.5mm]
& \llbracket \bPhi_L \rrbracket(\bxi,\omega)~=~\boldsymbol{a}(\bxi,\omega). \quad & \big(\bxi \in L, \omega \in \Omega \big)
\end{aligned}     
\vspace{-1.5mm}
\eeq
Given~\eqref{PhiL}, one may generate a library of physically-consistent scattering patterns on $S\obs$ for a grid of trial pairs $(\bx_\circ,\bR)$ sampling $\mathcal{S}\!\times\! U(3)$. 

The underpinning concept of wavefront shaping is that when the trial dislocation $L$ is a subset of the true scatterers $\Gamma$, its affiliated scattering pattern $\bPhi_L \in L^2(S\obs)^3 \nxs\times L^2(\Omega)^3$ may be recovered from experimental data by probing the range of operator $\Lambda$ i.e., through solving 
\vspace{-2 mm}
\beq\lb{FF}
\Lambda\exs \bg~\simeq~\bPhi_L, \qquad \bg \in L^2(S^{\text{inc}})^3\nxs\times L^2(\Omega)^3,
\vspace{-2 mm}
\eeq
for the wavefront densities $\bg(\bxi, \omega)$ on $\bxi \in S^{\textrm{inc}}$ at every frequency $\omega \in \Omega$.
In this setting, the principal theorem of linear sampling shines light on the unique behavior of $\bg$ in terms of $L$. This is accomplished by taking advantage of the factorization~\cite{Fatemeh2017,nguy2019}
 \vspace{-2mm}
\beq\lb{facts1}
\Lambda ~=~ \mathcal{H}^* \exs T \exs \mathcal{H}, 
\vspace{-2mm}
\eeq
where $()^*$ indicates the adjoint operator, and
\beq\nonumber 
\vspace{0mm}
\begin{array}{|ll}
\mathcal{H} \colon L^2(\partial\mathcal{B}_t)^3\nxs\times L^2(\Omega)^3 \rightarrow \exs H^{-\frac{1}{2}}(\Gamma)^3 \nxs\times L^2(\Omega)^3 & \\*[1mm]
\mathcal{H}(\bg) ~:=~ \bt^{\textrm{f}}(\bxi,\omega), \qquad \quad\,\, \big(\bxi \in \Gamma, \omega \in \Omega\big) & \\
\end{array} \quad \,\,
\begin{array}{|ll} 
\mathcal{H}^* \colon \tilde{H}^{\frac{1}{2}}(\Gamma)^3\nxs\times L^2(\Omega)^3 \rightarrow \exs L^2(\partial\mathcal{B}_t)^3 \nxs\times L^2(\Omega)^3 & \\*[1mm]
\mathcal{H}^*(\llbracket \bv \rrbracket) ~:=~ \bv(\bxi,\omega), \qquad  \big(\bxi \in \partial\mathcal{B}_t, \omega \in \Omega\big) & \\
\end{array}
\vspace{-1mm}
\eeq
\beq\lb{HTH}
\vspace{1.5mm}
\begin{array}{|ll}
T \colon H^{-\frac{1}{2}}(\Gamma)^3 \nxs\times L^2(\Omega)^3 \rightarrow \exs \tilde{H}^{\frac{1}{2}}(\Gamma)^3\nxs\times L^2(\Omega)^3 & \\*[1mm]
T(\bt^{\textrm{f}}) ~:=~ \llbracket \bv \rrbracket(\bxi,\omega). \qquad \quad \!\! \big(\bxi \in \Gamma, \omega \in \Omega\big) & \\
\end{array} 
\vspace{-1.5mm}
\eeq

This allows to rigorously characterize the solution $\bg$ according to~\cite[Theorem 6.2]{Fatemeh2017} as the following.
\begin{theorem}\lb{TR2}
Given~\eqref{facts1}, by assuming that the operator $\mathcal{H}(\Gamma,\omega)$ in~\eqref{HTH} is injective at frequency $\omega \in \Omega$,     
\begin{itemize}
\item~If $L\!\subset\!\Gamma$, there exists a density vector $\bg_\epsilon\!\in L^2(S^{\text{inc}})^3\!\times\nxs L^2(\Omega)^3$ such that $\|\Lambda\bg_\epsilon-\bPhi_L\|_{L^2(S\obs)} \leqslant\epsilon$ and $\limsup\limits_{\epsilon \rightarrow 0} \|\mathcal{H}\bg_\epsilon\|_{H^{-1/2}(\Gamma)}<\infty$.

\item~If $L \not\subset \Gamma$, then $\forall \bg_\epsilon\!\in L^2(S^{\text{inc}})^3\!\times\nxs L^2(\Omega)^3$ such that $\norms{\nxs \Lambda\bg_\epsilon-\bPhi_L  \nxs}_{L^2(S\obs)} \, \leqslant\epsilon$, $\,\lim\limits_{\epsilon \rightarrow 0} \norms{\mathcal{H}\bg_\epsilon}_{H^{-1/2}(\Gamma)} \,\,=\infty$. 
\end{itemize}
\end{theorem} 

\emph{LSM indicator.}~Theorem~\ref{TR2} of the linear sampling method poses two fundamental challenges in that:~(i) the featured anomaly indicator $\norms{\!\mathcal{H}\bg_\epsilon\!}_{H^{-1/2}(\Gamma)}$ inherently depends on the support of unknown scatterers $\Gamma$ since $\mathcal{H} = \mathcal{H}(\Gamma)$, and (ii) construction of the wavefront density $\bg_\epsilon \in L^2(S^{\text{inc}})^3\!\times\nxs L^2(\Omega)^3$ is implicit in the theorem~\cite{Audibert2014,Fatemeh2017}. Conventionally, these issues are addressed by replacing $\norms{\!\mathcal{H}\bg_\epsilon\!}_{H^{-1/2}(\Gamma)}\!$ with $\norms{\nxs\bg_\epsilon\!}_{L^2(S^{\text{inc}})}$ which is, in turn, computed by way of Tikhonov regularization
\vspace{-1.5mm}
\beq\lb{LSMg}
\textcolor{black}{
\bg_\epsilon \,\,=\,\, \bg_{\mathtt{L}}   \,\, \colon \!\!\! = \,\, \min_{\bg \in L^2(S^{\text{inc}})^3}  \big\lbrace \! \norms{\Lambda \bg \,-\, \bPhi_L}^2_{L^2(S\obs)} + \,\, \eta \! \norms{\bg}^2_{L^2(S^{\text{inc}})} \! \big\rbrace,} 
\vspace{-1.5mm}
\eeq
where~$\eta = \eta(L)>0$ is a regularization parameter computable by the Morozov discrepancy principle~\cite{Kress1999}.

On the basis of~\eqref{LSMg}, the LSM indicator functional for every frequency $\omega \in \Omega$ is constructed according to~\cite{Fiora2003} by 
\vspace{-1.5mm}
\beq\lb{LSM}
\mathtt{L} \,\, := \,\, \frac{1}{\norms{\bg_{\mathtt{L}}}_{L^2(S^{\text{inc}})}}.
\vspace{-0mm}
\eeq

$\mathtt{L} = \mathtt{L}(\bg_{\mathtt{L}},\omega)$ achieves its highest values at the loci of hidden scatterers $\Gamma$. More specifically, the behavior of $\mathtt{L}$ within the search volume $\mathcal{S}\subset\mathcal{B}$ may be characterized as the following, 
\vspace{-1.5mm}
\beq\lb{LSMB}
\begin{aligned}
& \text{if}\,\,\, L \subset \Gamma \quad \iff \quad  \liminf\limits_{\eta \rightarrow 0} \exs \mathtt{L}(\bg_{\mathtt{L}},\omega) \,>\, 0, \\*[-0.5mm]
& \text{if}\,\,\, L \subset \mathcal{S}\backslash \Gamma \quad \iff \quad \lim\limits_{\eta \rightarrow 0} \mathtt{L}(\bg_{\mathtt{L}},\omega) = 0. 
\end{aligned}
\vspace*{-1.5mm}
\eeq

\emph{GLSM indicator.}~Approximations underlying the LSM imaging functional may lead to instability of the reconstruction, and sensitivity to measurement errors (see Section~\ref{DI}).   To help meet the challenge, the GLSM~\cite{Fatemeh2017} takes advantage of the positive and self-adjoint operator $\Lambda_{\sharp}:\, L^2(\partial\mathcal{B}_t)^3\nxs\times L^2(\Omega)^3  \exs\to\exs L^2(\partial\mathcal{B}_t)^3\nxs\times L^2(\Omega)^3$, defined on the basis of the scattering operator $\Lambda$ by
\vspace{-1.5mm}
\beq\lb{Fsd}
\Lambda_{\sharp}\,\colon \!\!\!=\, \frac{1}{2} \big{|}\Lambda+\Lambda^*\big{|} \:+\: \big{|}\frac{1}{2 \textrm{\emph{i}}} (\Lambda\nxs-\Lambda^*)\big{|}, 
\vspace{-1.5mm}
\eeq
with the affiliated factorization~\cite{Kirsch2008} 
 \vspace{-1.5mm}
\beq\lb{facts2}
\Lambda_\sharp ~=~ \mathcal{H}^* \exs T_\sharp \exs \mathcal{H}, 
\vspace{-1mm}
\eeq
where the middle operator $T_\sharp$ is coercive according to~\cite[Lemma 5.7]{Fatemeh2017} i.e., there exists a constant $c>0$ independent of $\mathcal{H}\bg_\epsilon$ such that 
\vspace{-1mm}
\beq\lb{coT}
(\exs \bg_\epsilon, \exs \Lambda_\sharp \exs  \bg_\epsilon)_{L^2(\partial\mathcal{B}_t)^3} \, ~=~ \big \langle \mathcal{H}\bg_\epsilon, \, T_\sharp \mathcal{H}\bg_\epsilon \big \rangle_{\Gamma} ~\geqslant~ c \norms{\!\mathcal{H}\bg_\epsilon\!}^2_{H^{-1/2}(\Gamma)}, \qquad \forall  \mathcal{H}\bg_\epsilon \in H^{-1/2}(\Gamma)^3.
\vspace{-1mm}
\eeq

Here, $\dualGA{\cdot}{\cdot}$ denotes the duality product $\big\langle H^{-1/2}(\Gamma)^3, \tilde{H}^{1/2}(\Gamma)^3 \big\rangle$. Thanks to~\eqref{coT}, the term $\norms{\!\mathcal{H}\bg_\epsilon\!}^2_{H^{-1/2}(\Gamma)}$ in Theorem~\ref{TR2} may be safely replaced by $(\exs \bg_\epsilon, \exs \Lambda_\sharp \exs  \bg_\epsilon)_{L^2(\partial\mathcal{B}_t)^3}$ which is computable without prior knowledge of~$\Gamma$. Then, according to~\cite[Theorems 4.3]{pour2019} a robust solution to~\eqref{FF} may be constructed by
\vspace{-1.5mm}
\beq\lb{GLSMg}
\textcolor{black}{
\bg_\epsilon \,:=\,\, \min_{\bg \in L^2(S^{\text{inc}})^3}  \big\lbrace \! \norms{\Lambda \bg \,-\, \bPhi_L}^2_{L^2(S\obs)} + \,\, \gamma (\exs \bg, \exs \Lambda_\sharp \exs \bg)_{L^2(\partial\mathcal{B}_t)} \, + \, \delta \gamma \! \norms{\bg}^2_{L^2(S^{\text{inc}})} \! \big\rbrace,} 
\vspace{-1.5mm}
\eeq
without the heuristics involved in the LSM approach. It should be mentioned that, in~\eqref{GLSMg}, $\delta>0$ is a measure of noise in data, and $\gamma>0$ represents the regularization parameter defined in terms of $\eta(L)$ of~\eqref{LSMg} by 
\vspace{-1.5mm}
\beq\lb{Gamm}
\gamma(L) \,\, \colon \!\!\! = \,\, \frac{\eta(L)}{\norms{\Lambda}_{L^2} + \,\, \delta}.
\vspace{-1.5mm}
\eeq

Note that the GLSM cost functional~\eqref{GLSMg} is~\emph{convex}~\cite[Theorem 4.1]{pour2019}, and thus, its minimizer $\bg_\epsilon$ can be computed without iterations. Similar to the LSM indicator~\eqref{LSM}, the norm of penalty term in~\eqref{GLSMg} is used to identify the GLSM indicator as $[(\exs \bg_\epsilon, \exs \Lambda_\sharp \exs \bg_\epsilon)_{L^2(\partial\mathcal{B}_t)} \, + \, \delta \! \norms{\bg_\epsilon}^2_{L^2(S^{\text{inc}})}]^{-1/2}$.

\vspace{-1mm}
\begin{rem}[on the nature of $\Lambda_\sharp$]\label{ups}
The operator $\Lambda_\sharp$ is symmetric, and thus, amenable to specific sensing configurations where $S^{\text{inc}}\! = S\obs \subset \partial\mathcal{B}_t$. This implies that the loci of ultrasonic sources in experiments should coincide with the measurement points so that the discretized operator $\Lambda$ is a square matrix. This may not be plausible or efficient in practice as evidenced in section~\ref{exp_set} where the observation grid is ten times more dense than the excitation grid. 
\end{rem}
\vspace{-1mm}

This constraint may be relaxed by invoking Assumption~\ref{nodiss} where the system's energy dissipation is presumed negligible during the testing period $(0, T]$ so that the operator $\Lambda$ is normal~\cite{Kirsch2008}. In this setting,~\cite[Theorem 1.23]{Kirsch2008} indicates that there exists a second factorization 
\vspace{-1.5mm}
\beq\lb{facts2}
\Lambda ~=~ \big( \Lambda^* \Lambda \nxs\big)^{\!\frac{1}{4}} \, \text{\sf T} \exs \big( \Lambda^* \Lambda \nxs\big)^{\!\frac{1}{4}}, 
\vspace{-1.5mm}
\eeq 
such that the middle operator $\text{\sf T}$ is coercive, and the ranges of $\mathcal{H}^*$ in~\eqref{facts1} and $( \Lambda^* \Lambda)^{{1}/{4}}$ coincide. As a result, the term $\norms{\!\mathcal{H}\bg_\epsilon\!}^2_{H^{-1/2}(\Gamma)}$ in Theorem~\ref{TR2} may also be replaced by $(\exs \bg_\epsilon, \exs  ( \Lambda^* \Lambda)^{{1}/{2}} \exs  \bg_\epsilon)_{L^2(S^{\text{inc}})^3}$ which is computable from $\Lambda$ notwithstanding of its symmetry condition. Following~\cite[Theorems 4.3]{pour2019}, a solution to~\eqref{FF} is then generated by minimizing the modified GLSM cost functional, i.e., 
\vspace{-1.5mm}
\beq\lb{GLSMmg}
\bg_\epsilon \,\,=\,\, \bg_{\mathtt{G}}   \,\, \colon \!\!\! =  \min_{\bg \in L^2(S^{\text{inc}})^3}  \big\lbrace \! \norms{\Lambda \bg \,-\, \bPhi_L}^2_{L^2(S\obs)} + \,\, \gamma (\exs \bg, \exs \Upsilon \exs \bg)_{L^2(S^{\text{inc}})} \, + \, \delta \gamma \! \norms{\bg}^2_{L^2(S^{\text{inc}})} \! \big\rbrace, \quad  \Upsilon =  \big( \Lambda^* \Lambda \big)^{\!\frac{1}{2}},
\vspace{-1.5mm}
\eeq

The new cost functional~\eqref{GLSMmg} is also convex and its minimizer $\bg_{\mathtt{G}} = \bg_{\mathtt{G}}(L,\omega)$ may be obtained non-iteratively as elucidated in section~\ref{DI}. Following~\cite[Theorems 4.3]{pour2019}, one may show that as $\gamma \rightarrow 0$, the solution $\bg_{\mathtt{G}}$ remains bounded if and only if $L \subset \Gamma$. More specifically,~at every frequency $\omega \in \Omega$,
\vspace{-1.5mm}
\beq\lb{statG1}
\begin{aligned}
& \text{if}\,\,\, L \subset \Gamma ~\iff~ \limsup\limits_{\gamma \rightarrow 0}\limsup\limits_{\delta \rightarrow 0} \Big( \nxs (\exs \bg_{\mathtt{G}}, \Upsilon\bg_{\mathtt{G}})_{L^2(S^{\text{inc}})} \!\,+\, \delta  \norms{\nxs \bg_{\mathtt{G}} \nxs}^2_{L^2(S^{\text{inc}})}  \!\!\Big) \,<\, \infty, \\
&\text{if}\,\,\, L \not\subset \Gamma ~\iff~ \liminf\limits_{\gamma \rightarrow 0}\liminf\limits_{\delta \rightarrow 0} \Big( \nxs (\exs \bg_{\mathtt{G}}, \Upsilon \bg_{\mathtt{G}})_{L^2(S^{\text{inc}})} \!\,+\, \delta  \norms{\nxs \bg_{\mathtt{G}} \nxs}^2_{L^2(S^{\text{inc}})}  \!\!\Big) \,=\, \infty.
 \end{aligned}
\vspace{-1.5mm}
\eeq

Based on this, the (modified) GLSM indicator functional is defined by  
\vspace{-1.5mm}
\beq\lb{EIFn0}
\mathtt{G}(\bg_{\mathtt{G}},\omega) \,\, := \frac{1}{\sqrt{(\exs \bg_{\mathtt{G}}, \Upsilon \bg_{\mathtt{G}})_{L^2(S^{\text{inc}})} \, + \, \delta \! \norms{\bg_{\mathtt{G}}}^2_{L^2(S^{\text{inc}})}}},  
\vspace*{-1.5mm}
\eeq
which reconstructs the support of hidden scatterers by achieving its highest values near $\Gamma$. 
\vspace*{-3mm}
\section{Experimental campaign} \label{exp_set}

\noindent Experiments are performed on a prismatic specimen of charcoal granite of dimensions $0.96$m $\!\!\times\exs 0.3$m $\!\!\times\exs 0.03$m, mass density $\rho\!=\!2750$kg/m$^3$, nominal Poisson's ratio $\nu\!=\!0.23$, and nominal Young's modulus $E\!=\!62.6$GPa. These values are identified via a uniaxial compression test on a cylindrical sample of the same material. 

The testing procedure involves three steps:~(i) elastic-wave excitation and sensing in the baseline system,~(ii)~fracturing of the specimen, and (iii) elastic-wave testing of the damaged system.

\begin{figure}[!tp] \vspace*{-0mm}
\center\includegraphics[width=0.85\linewidth]{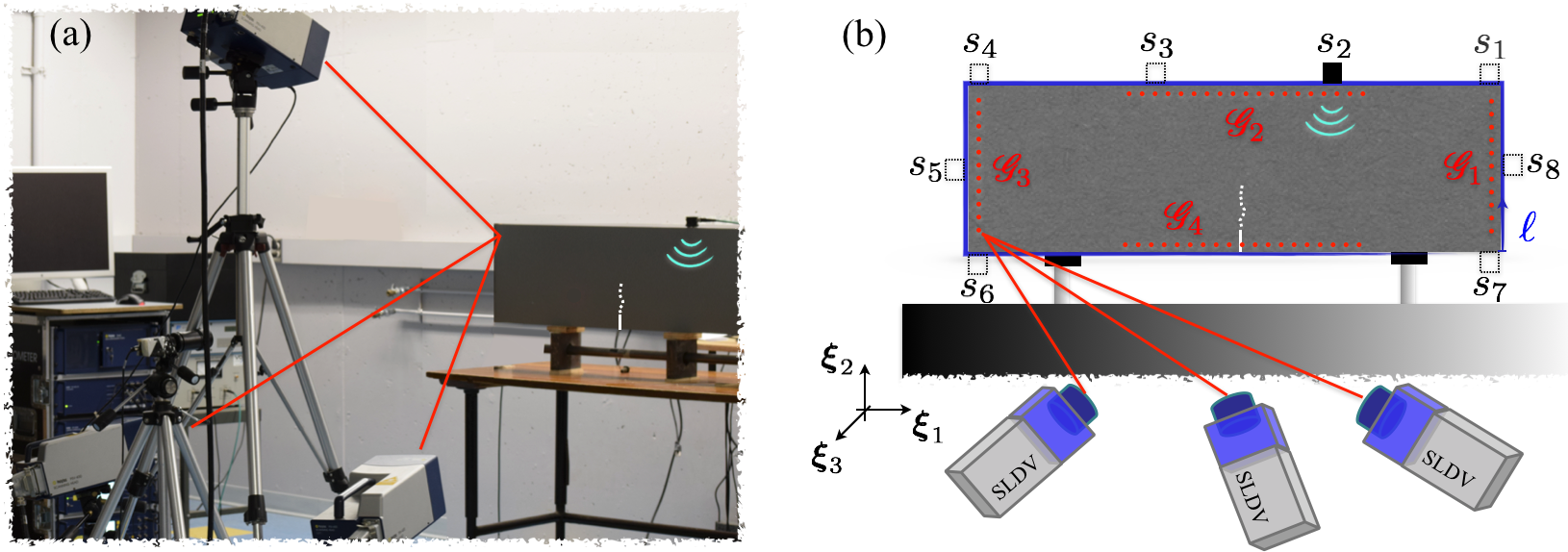} \vspace{-3.5mm}
\caption{Testing set-up:~(a)~a prismatic slab of charcoal granite is subject to ultrasonic testing prior to and after being fractured via three-point bending;~(b)~shear waves are generated by a piezoelectric source at~$s_i$ \mbox{$(i\!=\!1,2,\ldots,8)$}, and the triaxial particle velocity field is captured by~a 3D SLDV over the designated scanning grid~$\bigcup_{i=1}^4\mathcal{G}_i$.} \label{Exp-sch} \vspace{-4mm}
\end{figure}

\emph{Step~1.} The ultrasonic experiments are first performed on the intact granite slab as shown in Fig.~\ref{Exp-sch} (a). Waveforms measured in this step furnish the ``baseline'' response of the system associated with the incident field $\bu\ff(\bxi,t)$. This is required for computing the scattered field $\bv(\bxi,t) = \bu(\bxi,t) - \bu\ff(\bxi,t)$, wherein $\bu(\bxi,t)$ represents the \emph{total field} measurements in \emph{Step 3}. \emph{Step 1} entails eight ultrasonic experiments where the sample is excited by an \emph{in-plane} shear wave from one of the designated source locations $s_1, s_2, \ldots, s_{8}$ shown in Fig.~\ref{Exp-sch}(b). Shear waves are generated by a 0.5 MHz piezoelectric transducer (V151-RB by Olympus, Inc.)~whose diameter of 32 mm is almost commensurate with the granite thickness. The transducer is aligned with the granite mid-plane along $\bxi_3$ minimizing the out-of-plane excitation. The incident signal is a five-cycle burst of the form 
\vspace{-1.5mm}
\begin{equation}\label{wavelet}
H({\sf f_c}t) \, H(5\!-\!{\sf f_c}t) \, \sin\big(0.2 \pi {\sf f_c} t\big) \, \sin\big(2 \pi {\sf f_c} t\big), 
\vspace{-1.5mm}
\end{equation}
where ${\sf f_c}\!=\!30\mbox{kHz}$ denotes the center frequency, and $H$ is the Heaviside step function. The induced wave motion from each source location is measured by a 3D Scanning Laser Doppler Vibrometer (SLDV) as shown in Fig.~\ref{Exp-sch}(a). The PSV-400-3D SLDV system by Polytec, Inc.~is capable of capturing the triaxial components of particle velocity on the surface of solids over a designated scanning grid. Its measurement (\emph{resp.}~spatial) resolution is better than 1$\mu${m}/s (\emph{resp.}~0.1mm) within the frequency range DC-1MHz, facilitating waveform sensing in the nanometer scale in terms of displacement~\citep{Polytec}.  

\emph{Step~2.} A notch of length $4$cm and width $1.5$mm is manufactured at the bottom center of specimen. The sample is then fractured in the three-point-bending (3PB) configuration by a closed-loop, servo-hydraulic, $1000$kN MTS load frame such that the crack propagation is controlled by the crack mouth opening displacement (CMOD) measured by a clip gage. The loading process is monotonic with respect to the CMOD at a constant rate of $0.1\mu$m/s. The loading process is continued up to approximately $65\%$ of the maximum force in the post-peak regime with the associated CMOD of $320\mu$m. Upon completion of the fracturing process, the specimen is unloaded and reconfigured according to Fig.~\ref{Exp-sch} (a).  

\emph{Step~3.} The ultrasonic experiments are performed on the fractured specimen following the same procedure as in \emph{Step~1}, i.e., the testing set-up involving the transducer locations, illuminating wavelet, and scanning area is as shown in Fig.~\ref{Exp-sch}.

\vspace{-1mm}
\begin{rem}[on the nature of wave motion]\label{pw}
Measurements may be interpreted in the context of \emph{plane stress} approximation -- related to the elastic analysis of thin plates~\citep{Mal1969}, whereby the particle motion is considered invariant through the thickness of specimen. In this setting, the \emph{effective} Poisson's ratio and Young's modulus are respectively identified by $\nu'\!=\!\nu/(1+\nu)$ and $E'\!=\!E(1-\nu'^2)$~\citep{Mal1969}, resulting in the shear (S-) and compressional (P-) wave velocities 
\vspace{-1.75mm}
\begin{equation}\label{cps}
c_s ~=~ \sqrt{\frac{E}{2(1+\nu)\rho}} ~=~ 3041 \, \, \mbox{m/s}, \qquad  c_p ~=~ \sqrt{\frac{E}{(1-\nu^2)\rho}} ~=~ 4901 \, \, \mbox{m/s}. 
\vspace{-1.25mm}
\end{equation}
Observe that the shear wavelength $\lambda_s$ in the specimen may be approximated by $10$cm at $30\mbox{kHz}$, giving the shear-wavelenghth-to-plate-thickness ratio of $\lambda_s/h\nxs\gtrsim\nxs3.3$. In this range, the phase error committed by the plane stress approximation is about $3\%$~\citep{Lamb1917}. An in-depth experimental analysis of plane-stress wave propagation -- in a specimen of similar dimensions and material properties, is provided in~\cite{pour2018} where full-field waveform data are analyzed within the frequency range $10\nxs-\nxs40\mbox{kHz}$. 

It should be mentioned that the sampling approaches to inverse scattering are full-waveform inversions~\cite{pour2019}, and thus, they do not rely on a specific mode of propagation, nor they require any such knowledge on the nature of wave motion. In this study, the plane-stress approximation implies that the data inversion may be conducted in a reduced-order space involving the in-plane components of the measured response as delineated in section~\ref{DI}.      
\end{rem}

\vspace{-1mm}
As illustrated in Fig.~\ref{Exp-sch}(b), the scanning grid $\bigcup_{i=1}^4\mathcal{G}_i$ is in the immediate vicinity of the external boundary of specimen. More specifically, $\mathcal{G}_1$ (\emph{resp.}~$\mathcal{G}_3$) is centered in the mid- right (\emph{resp.}~left) edge of the sample with $27$ uniformly spaced measurement points over a span of $22$cm, while $\mathcal{G}_2$ (\emph{resp.}~$\mathcal{G}_4$) is at the top (\emph{resp.}~bottom) center of the plate involving a uniform grid of $45$ scan points over an interval of $38$cm. In light of Remark~\ref{pw}, this amounts to a spatial resolution of about 8mm for ultrasonic measurements at $30\mbox{kHz}$ in $\bxi_1\nxs$ and $\bxi_2$ directions. At every scan point, the data acquisition is conducted for a time period of 1ms at the sampling rate of 512kHz. To minimize the impact of (optical and mechanical) random noise in the system, the measurements are averaged over an ensemble of 60 realizations at each scan point. Furthermore, signal enhancement and speckle tracking were enabled to avoid signal dropouts due to surface roughness.       

\vspace{-1mm}
\begin{rem}\label{sg}
Note that the observation grid is consistent with common configurations in practice where only a subset of the domain's external boundary is accessible for (contact or non-contact) sensing. Recall that the (G)LSM indicators reconstruct the support of internal scatterers from boundary data. Thus, full-field ultrasonic measurements i.e., waveforms on the entire surface of specimen are not captured in this study. An image processing scheme for anomaly detection by way of full-field measurements is provided in~\cite{pour2018}.  
\end{rem}
\vspace{-1mm}

\begin{figure}[!tp] \vspace*{-0mm}
\center\includegraphics[width=0.98\linewidth]{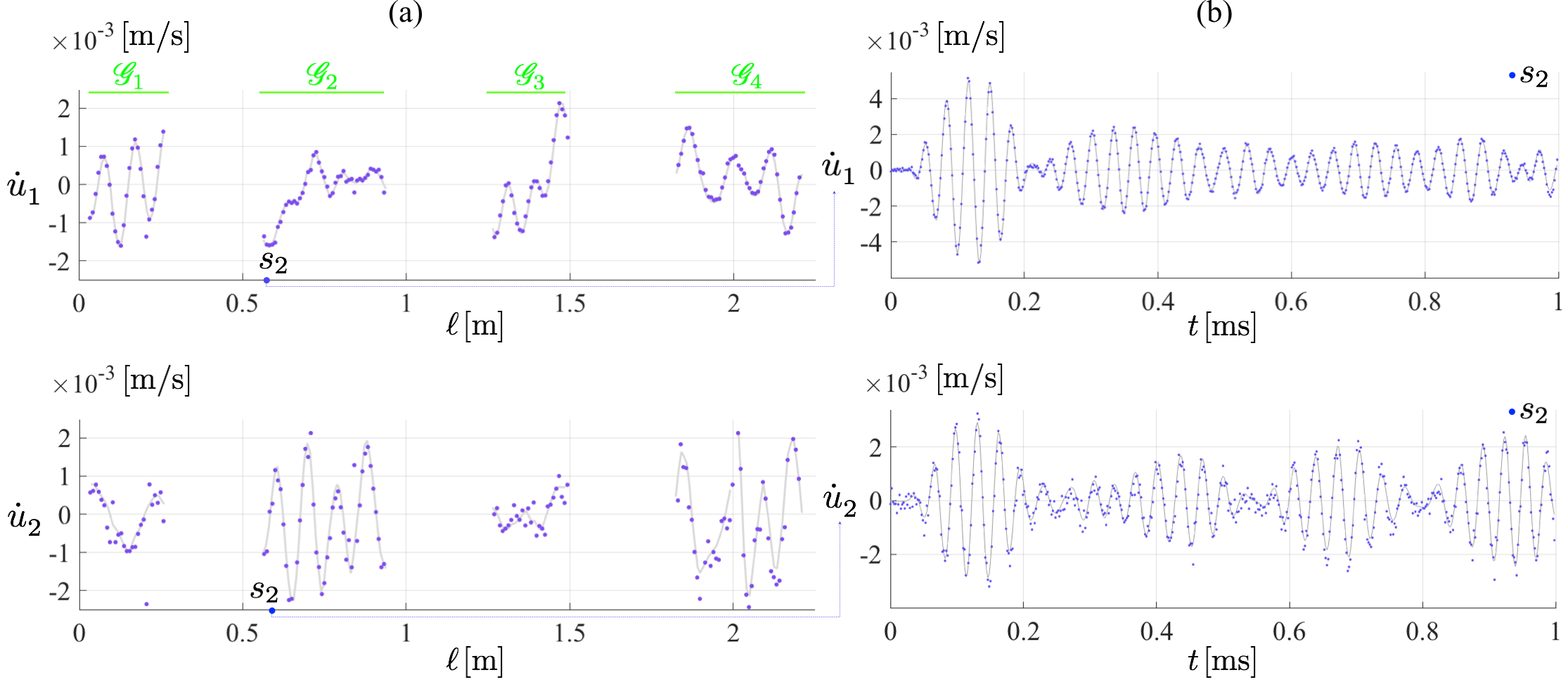} \vspace{-3.5mm}
\caption{SLDV measurements over the scanning grid $\bigcup_{i=1}^4\mathcal{G}_i$:~(a)~particle velocity distribution $\dot{u}_1(\ell,t=0.25\textrm{ms})$ (\emph{resp.}~$\dot{u}_2(\ell,t=0.25\textrm{ms})$) in $\bxi_1$ (\emph{resp.}~$\bxi_2$) direction, where $\ell$ represents the counterclockwise arc length along the specimen's edge as in Fig.~\ref{Exp-sch}(b),~and~(b)~time history of the particle velocity response $[\dot{u}_1 \,\, \dot{u}_2](\ell=0.6\textrm{m},t)$ measured in the vicinity of transducer at $s_2$. Dots represent ``raw" measurements and solid lines are the corresponding processed data according to section~\ref{DSP}.} \label{RS} \vspace{-5mm}
\end{figure}

To demonstrate the acquired SLDV measurements, Fig.~\ref{RS}(a) displays a snapshot in time (at $t = 0.25$ms) of the particle velocity distributions $\dot{u}_1$ and $\dot{u}_2$ over the scanning grid $\bigcup_{i=1}^4\mathcal{G}_i$ in $\bxi_1$ and $\bxi_2$ directions, respectively. These measurements are conducted on the intact specimen prior to fracturing. Note that the test data is plotted against the counterclockwise arc length $\ell$ around the specimen's external boundary whose origin is at the bottom-right corner of the plate as shown in Fig.~\ref{Exp-sch}(b). Fig.~\ref{RS}(b) plots the time history of in-plane SLDV measurements at a fixed grid point with the affiliated arc length $\ell = 0.6$m -- in the immediate vicinity of the ultrasonic source~$s_2$ indicated in Fig.~\ref{RS}(a). It should be mentioned that in Fig.~\ref{RS}, ``raw" test data are shown with dots, while the processed data (according to section~\ref{DSP}) are shown by the linearly interpolated solid lines.  

\vspace{-1mm}
\begin{rem}[scattered field data]\label{sw}
\textcolor{black}{Recall that the (G)LSM indicators rely on the spectrum of scattered field $\bv$ which may be directly computed from the free field $\bu\ff$ measured in \emph{Step 1}, and total field $\bu$ captured in \emph{Step 3}. An effort was made to generate sufficiently similar incident waveforms (up to some simple post processing measures described in section~\ref{DSP}) at each source location in both sensing steps. This is accomplished by exercising:~(i)~precise geometric alignment of the piezoelectric transducer, (ii)~application of a thin and uniform layer of cyanoacrylate glue as couplant, and (iii)~comparison of the incident waveforms captured in the vicinity of the transducer (before any reflections occur) prior to conducting the planned data acquisition.}
\end{rem}
\vspace{-1mm}
 
\vspace{-2.5mm}
\section{Signal processing}\label{DSP}

\noindent  This section aims to systemically extract the spectrum of scattered displacement response over the observation grid from the SLDV-measured particle velocity data. The results will be deployed in section~\ref{DI} to reconstruct the support of 3PB-induced damage in the granite specimen. In this vein, ``raw" measurement data are processed in three stages, involving:~(1)~spatiotemporal filtering and time integration,~(2)~synchronization of incidents and extraction of scattered fields, and~(3)~spectral analysis. 

{\slshape (1)~spatiotemporal filtering and time integration.} A band-pass filter of bandwidth $20$kHz centered at $30$kHz -- consistent with the spectrum of excitation wavelet~(\ref{wavelet}), is applied to the particle-velocity records at every scan point. Note that the filtered velocity signals are temporally smooth and differentiable as shown by solid lines in Fig.~\ref{RS}(b). At every snapshot in time, however, the spatial distribution of particle velocity over the scanning grid is contaminated with data points of exceptionally low signal-to-noise ratio -- identified by sudden spikes in the observed waveforms e.g.,~see Fig.~\ref{RS}(a). To mitigate the spatial noise, first, a \emph{unified} set of observation points are specified on $\bigcup_{i=1}^4\mathcal{G}_i$ which remain invariant for both datasets obtained in \emph{Steps 1 and 3} of the testing procedure (prior to and after fracturing the specimen). Then, at every time sample, four linear interpolation functions are constructed independently on $\mathcal{G}_1, \ldots, \mathcal{G}_4$ making use of (temporally filtered) velocity data points of admissible signal-to-noise ratio i.e.,~noisy points are excluded from the interpolation. In this setting, the velocity distribution at a given time may be computed over the unified observation points via the indicated interpolants. The resulting waveforms are spatially smooth as shown by solid lines in Fig.~\ref{RS}(a). A unified observation grid enables arithmetic operations between datasets of distinct sensing steps 1 and 3, which is required for the computation of scattered field. Thus-obtained velocity signals are then transformed into displacement data through numerical integration. The latter process, however, introduces a low-frequency drift i.e.,~integration constant in the results, which is eliminated by a high-pass filter of cut-off frequency $500$Hz. In this way, one finds the spatiotemporally smooth ``total" displacement fields corresponding to $\bu(\bxi,t)$ in \eqref{uk} over $S\obs$ which calls for further processing since the ``scattered"  field $\bv(\bxi,t)$ will be needed for the reconstructions of section~\ref{DI}.  

\begin{figure}[!tp] \vspace*{-0mm}
\center\includegraphics[width=0.98\linewidth]{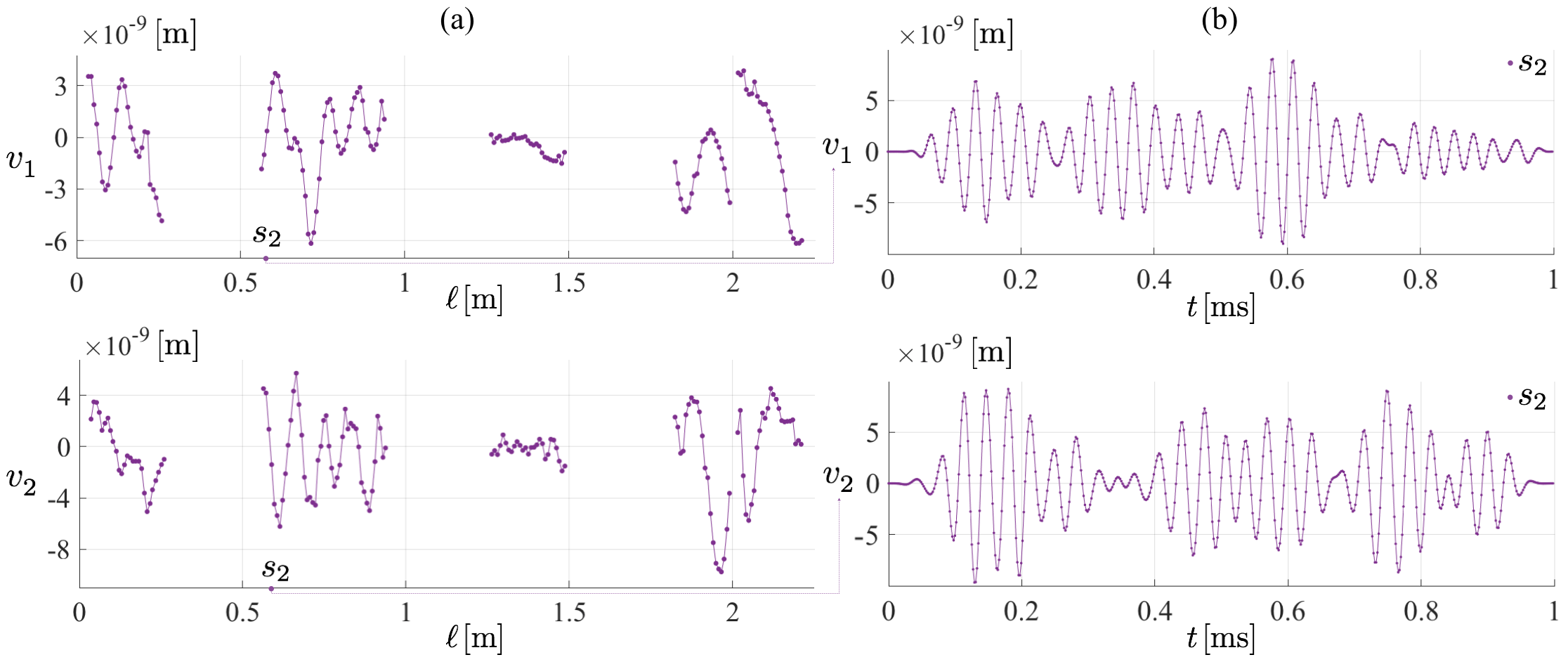} \vspace{-3.5mm}
\caption{Spatiotemporal scattered displacement field:~(a)~in-plane displacement distributions ${v}_1(\ell,t=0.25\textrm{ms})$ and ${v}_2(\ell,t=0.25\textrm{ms})$ -- in $\bxi_1\nxs$ and $\bxi_2$ directions, where $\ell$ is the arc length, and~(b)~time history of the scattered displacement response $[{v}_1 \,\, {v}_2](\ell=0.6\textrm{m},t)$ computed in the vicinity of the ultrasonic source at~$s_2$. Dots are the response affiliated with the \emph{unified} observation points, while the solid lines linearly interpolate the data points to clarify the waveforms.} \label{PS} \vspace{-4mm}
\end{figure}

{\slshape (2)~synchronization of incidents and extraction of scattered fields.}~To calculate the scattered field in light of remark~\ref{sw}, this stage aims to synchronize the time, and balance the magnitude of ultrasonic incidents between \emph{Steps 1 and 3} of experiments. Discrepancies in transducer's physical input at different sensing steps -- although curtailed by the measures indicated in the remark, are inevitable due to~(a)~perturbation of transducer-specimen coupling in reattachments, and~(b)~recalibration of the 3D SLDV system for ultrasonic tests of \emph{Step 3} (after fracturing the specimen). To address this problem, let us consider the (processed) incident displacement fields $\bu\ff$ (related to the intact specimen) in the vicinity of every ultrasonic source $s_1, \ldots, s_8$. The support of which is a subset of:~(a)~$\mathcal{G}_1$ near $s_1$, $s_8$, and $s_7$,~(b)~$\mathcal{G}_2$ in the immediate vicinity of $s_2$ and $s_3$, and~(c)~$\mathcal{G}_3$ in a neighborhood of $s_4$, $s_5$, and $s_6$. Then, the ``reference" physical incidents (transducer inputs) are identified as the first 80-100 samples of displacement time histories in the indicated neighborhoods of $s_1, \ldots, s_8$. Note that within this timeframe i.e.,~[0 0.15]ms to [0 0.2]ms depending on the source location, there is no fingerprint on the measured waveforms due to internal scatterers. In this setting, the displacement fields from every ultrasonic experiment in \emph{Step 3} are uniformly scaled (by a constant value) and shifted in time (by a fixed amount) so that the transducer inputs in \emph{Step 3} matches their counterparts in \emph{Step 1} for every source location. This leads to consistent ultrasonic data for both sensing steps, and one may now proceed to compute the scattered displacement fields by subtracting the \emph{total} fields from their associated incidents fields. Fig.~\ref{PS} illustrates the resulting scattered field distribution in time and space when the transducer is at $s_2$.   

{\slshape (3)~spectral analysis.}~This stage computes the spectrum of scattered displacement signals obtained in stage (2). Prior to the application of discrete Fourier transform, the problem of ``spectral leakage"~\cite{ewin1984} due to the transient nature of measured waveforms should be addressed. In this vein, the displacement data are temporally windowed~\cite{Oppe1999} using a tapered cosine i.e.~Tukey window of the form~\cite{bloo2004},   
\vspace{-1.5 mm}
\beq\label{tuw}\nonumber
\bw(\text{\bf t},\text{\sf c}) \,=\,
\!\left\{\begin{array}{l}
\vspace{0.5mm}
\!\! \dfrac{1}{2} \Big[1+\cos\nxs\big(\dfrac{2\pi}{\text{\sf c}T}(\text{\bf t}-{\text{\sf c} T}/{2})\big)\nxs\Big], \hspace{13.5mm} 0 \exs\leqslant\exs \text{\bf t} \exs<\exs  \dfrac{\text{\sf c} T}{2} \!\!\! \\*[0.5mm]
\!1, \hspace{52mm} \dfrac{\text{\sf c} T}{2} \exs\leqslant\exs \text{\bf t} \exs<\exs T-\dfrac{\text{\sf c} T}{2} \!\!\! \\*[2mm]
\!\! \dfrac{1}{2} \Big[1+\cos\nxs\big(\dfrac{2\pi}{\text{\sf c}T}(\text{\bf t}-T\!+{\text{\sf c} T}/{2})\big)\nxs\Big], \qquad T-\dfrac{\text{\sf c} T}{2} \exs\leqslant\exs \text{\bf t} \exs\leqslant\exs  T \!\!\!
\end{array}\right.
\vspace{-1.5 mm}
\eeq
where $T$ signifies the observation interval $[0 \,\,\, 1]$ms; $\text{\bf t}$ is the sampled time vector of length 512, and $0 \leqslant \text{\sf c} \leqslant 1$ is the ratio of cosine-tapered length to the entire window length. Fig.~\ref{PS}(b) shows the scattered displacements at $s_2$ after the application of Tukey window $\bw(\text{\bf t},0.2)$. Now that the support of windowed time signals is compact, one may safely proceed to compute the spectrum of scattered displacement fields via the fast Fourier transform. The resulting waveforms in the frequency domain will be used for the reconstruction in section~\ref{DI}.

\vspace{-1.5mm}
\section{Data Inversion}\label{DI}

With the preceding data, one may generate the (G)LSM indicator maps in three steps, namely by:~(i) constructing the discrete scattering operator ${\boldsymbol{\Lambda}}$, (ii) computing the trial signature patterns affiliated with~\eqref{PhiL}, and (iii) evaluating the (G)LSM imaging functionals~\eqref{LSM}~and~\eqref{EIFn0} through non-iterative minimization of their corresponding cost functionals~\eqref{LSMg} and~\eqref{GLSMmg}. These steps are elucidated in the following. 

\vspace{-1.5mm}
\subsection{The discrete scattering operator}
With reference to Fig.~\ref{Exp-sch}(b), the incident surface $S^{\text{inc}}$ is sampled at $N_s = 8$ source locations $\by_j \in \lbrace s_1, s_2, \ldots, s_{8} \rbrace$, while the observation grid $S\obs = \bigcup_{\iota=1}^4\mathcal{G}_\iota$ is comprised of $N_p = 144$ measurement points $\bxi_i$. In this setting, the spectrum of (in-plane) waveform data at $N_\omega = 10$ frequencies, specifically at~$\omega_\ell = 27, 28, \ldots, 36$kHz, are deployed to generate the multi-frequency scattering operator ${\boldsymbol{\Lambda}}$ as a $2N_pN_\omega\!\times N_sN_\omega$ matrix of components     
\vspace{-1.5 mm}
\beq\lb{mat2}
{\boldsymbol{\Lambda}}(2N_p\ell+2i+1\!:\!2N_p\ell+2i+2, \,N_s\ell+ j+1) ~=\, 
\left[\begin{array}{c}
\!\!F(v_1)\!\!   \\*[1mm]
\!\!F(v_2)\!\!  
\end{array}\!\right] \! (\bxi_i,\by_j;\omega_\ell),
\vspace{-1.5 mm}
\eeq
for 
\vspace{-1.5 mm}
\beq\lb{ijl}
i = 0,\ldots N_p-1, \quad j = 0,\ldots N_s-1,  \quad \ell = 0,\ldots N_\omega-1.
\vspace{-1.5 mm}
\eeq
On recalling~\eqref{So}, here, $F(v_\iota)(\bxi_i,\by_j;\omega_\ell)$, $\iota = 1,2$, is the $\iota^{\textrm{th}}\nxs$ component of the Fourier transformed displacement at the observation point $\bxi_i$ and frequency $\omega_\ell$ when the ultrasonic source is located at $\by_j$. 

\vspace{-1.5mm}
\subsection{A physics-based library of trial patterns}
Let the search volume $\mathcal{S}$ be a $29$cm $\nxs\!\times\!\nxs$ $29$cm square in the middle of specimen probed by a uniform $100 \!\times\! 100$ grid of sampling points~$\bx_{\small \circ}$ where the featured (G)LSM indicator functionals~\eqref{LSM}~and~\eqref{EIFn0} are evaluated. In addition, the unit circle is sampled by $16$ trial normal directions $\textrm{\bf{n}}=\bR\bn_{\small \circ}$ wherein $\bn_{\small \circ}=(1,0)$. Based on this, a total of $M = 10000 \!\times\! 16$ trial dislocations $L =\bx_\circ\!+\bR{\sf L}$ are generated for the specified pairs $(\bx_{\small \circ},\textrm{\bf{n}})$. Here, ${\sf L}$ is a vertical crack of length 3mm. For each $(\bx_{\small \circ},\textrm{\bf{n}})$, the scattering signatures $\textrm{\bf{v}}^{\bx_{\small \circ\nxs},\textrm{\bf{n}}}(\bxi_i,\omega)$ are computed separately for every $\omega \in \Omega := \lbrace 27, 28, \ldots, 36 \rbrace$kHz over the observation grid $\bxi_i \in S^{\text{obs}}$ by solving
\vspace{-1.5mm}
\beq\lb{PhiL2}
\begin{aligned}
&\nabla \nxs\cdot [\bC \exs \colon \! \nabla \textrm{\bf{v}}^{\bx_{\small \circ\nxs},\textrm{\bf{n}}}](\bxi,\omega) \,+\, \rho \exs \omega^2\textrm{\bf{v}}^{\bx_{\small \circ\nxs},\textrm{\bf{n}}}(\bxi,\omega)~=~\bzero, \quad & \big(\bxi \in {\mathcal{B}}\backslash L, \omega \in \Omega \big) \\*[0.5mm]
&\bn \nxs\cdot \bC \exs \colon \!  \nabla  \textrm{\bf{v}}^{\bx_{\small \circ\nxs},\textrm{\bf{n}}}(\bxi,\omega)~=~\bzero,  \quad & \big(\bxi \in \partial{\mathcal{B}}\backslash S, \omega \in \Omega \big) \\*[0.5mm]
&\textrm{\bf{v}}^{\bx_{\small \circ\nxs},\textrm{\bf{n}}}(\bxi,\omega)~=~\bzero,  \quad & \big(\bxi \in S, \omega \in \Omega \big) \\*[0.5mm]
& \textrm{\bf{n}} \cdot \bC \exs \colon \!  \nabla  \textrm{\bf{v}}^{\bx_{\small \circ\nxs},\textrm{\bf{n}}} ~=~ |{\sf L}|^{-1} \delta (\bxi-\bx_{\small \circ}\!) \exs \textrm{\bf{n}}. \quad & \big(\bxi \in L, \omega \in \Omega \big) \\
\end{aligned}     
\vspace{-0.0mm}
\eeq

Here, $\mathcal{B}$ represents the granite specimen, and $S$ represents the 2-cm long contact areas at the bottom of the plate where the wood supports meet the sample as shown in Fig.~\ref{Exp-sch} (a). 

These simulations are performed in three dimensions for the $0.96$m $\!\!\times\exs 0.3$m $\!\!\times\exs 0.03$m granite plate via an elastodynamics code rooted in the boundary element method~\cite{Bon1999, Fatemeh2015}. For data inversion, however, only the in-plane components of the computed scattered fields are used in the following form 
\vspace{-1.5 mm}
\beq\lb{Phi-inf-Dnum}
\bPhi_{\bx_{\small \circ},\textrm{\bf{n}}}(2N_p\ell+2i+1\!:\!2N_p\ell+2i+2) ~=~ 
\! \left[\begin{array}{c}  \!\! \textrm{{v}}_1^{\bx_{\small \circ\nxs},\textrm{\bf{n}}} \!\!\nxs \\*[1mm]
\!\! \textrm{{v}}_2^{\bx_{\small \circ\nxs},\textrm{\bf{n}}} \!\!\nxs
\end{array} \right]\!\nxs (\bxi_i;\omega_\ell), \qquad  i = 0,\ldots N_p-1,  \quad \ell = 0,\ldots N_\omega-1,
\vspace{-1.5 mm}
\eeq
where $\bPhi_{\bx_{\small \circ},\textrm{\bf{n}}}$ is a $2N_pN_\omega\!\times\! 1$ vector. In this setting, the scattering equation~\eqref{FF} may be discretized as
\vspace{-1.5 mm}
\beq\lb{Dff}
{\boldsymbol{\Lambda}} \, \bg_{\bx_{\small \circ\nxs},\textrm{\bf{n}}}~=~\bPhi_{\bx_{\small \circ},\textrm{\bf{n}}}. 
\vspace{-1.5 mm}
\eeq 

 \begin{rem}
 It is worth noting that $\bPhi_{\bx_{\small \circ},\textrm{\bf{n}}}$ is invariant with respect to ${\boldsymbol{\Lambda}}$. Hence, for computational efficiency, one may generate a $2N_pN_\omega \!\times\! M$ matrix $\bPhi$,
 \vspace{-1.5 mm}
\beq\lb{Phi-inf-Dnum2}\nonumber
\bPhi(2N_p\ell+2i+1\!:\!2N_p\ell+2i+2,m) ~=~ 
\! \left[\begin{array}{c}  \!\! \textrm{\emph{v}}_1^{(\bx_{\small \circ\nxs},\textrm{\bf{n}})_m} \!\!\! \\*[1mm]
\!\! \textrm{\emph{v}}_2^{(\bx_{\small \circ\nxs},\textrm{\bf{n}})_m} \!\!\!
\end{array} \right]\!\! (\bxi_i;\omega_\ell), \qquad  i = 0,\ldots N_p-1,  \quad \ell = 0,\ldots N_\omega-1,
\vspace{-1.5 mm}
\eeq
as the right hand side of scattering equation (\ref{Dff}) -- encompassing all choices of trial pairs $(\bx_{\small \circ},\textrm{\bf{n}})_m$, $m = 1, 2, \ldots M$.       
\vspace{-1.5mm}
 \end{rem}

\vspace{-1.5mm}
\subsection{The (generalized) linear sampling indicators}
The scattering equation~(\ref{Dff}) is generally ill-posed due to~{(a)}~nonlinear nature of the inverse problem,~{(b)}~limited excitation and sensing apertures,~{(c)}~local (e.g., interfacial) modes of wave motion whose signature may not be found on~$S\obs$~\cite{Fatemeh2017(2)}, and~{(d)}~noise in data. Accordingly, (\ref{Dff}) is primarily solved by regularization e.g.,~through minimizing a designated (Tikhonov or GLSM) cost functional, or via sparse sampling.  

\vspace{-1.5mm}
\subsubsection{The classical linear sampling indicator}
Following~\cite{Fatemeh2017}, the Tikhonov-regularized solution $\bg^\mathfrak{T}_{\bx_{\small \circ\nxs},\textrm{\bf{n}}}$ to (\ref{Dff}) is computed by non-iteratively minimizing the LSM cost functional,  
\vspace{-1 mm}  
\beq\label{lssm1}
\bg^\mathfrak{T}_{\bx_{\small \circ\nxs},\textrm{\bf{n}}} \,\,\colon \!\!= \,\, \text{argmin}_{\bg_{\bx_{\small \circ\nxs},\textrm{\bf{n}}}} \Big\{  \norms{{\boldsymbol{\Lambda}} \, \bg_{\bx_{\small \circ\nxs},\textrm{\bf{n}}}\,-\,\bPhi_{\bx_{\small \circ},\textrm{\bf{n}}}}^2_{L^2} \,+\,\,\,  \eta_{\bx_{\small \circ\nxs},\textrm{\bf{n}}} \norms{\bg_{\bx_{\small \circ\nxs},\textrm{\bf{n}}}}^2_{L^2}\Big\},
\vspace{-1 mm}  
\eeq
where the regularization parameter $\eta_{\bx_{\small \circ\nxs},\textrm{\bf{n}}}$ is obtained by way of Morozov discrepancy principle~\cite{Kress1999}. On the basis of~\eqref{lssm1}, the LSM indicator functional is constructed as 
\vspace{-1 mm}  
\beq\lb{LSM2}
\mathfrak{L_T}(\bx_{\small \circ\nxs}) \,\, = \,\, \frac{1}{\norms{\bg^{\mathfrak{T}}_{\bx_{\small \circ\nxs}}\!}_{L^2}}, \qquad
\textcolor{black}{
\bg^{\mathfrak{T}}_{\bx_{\small \circ\nxs}} \,\,\colon \!\!= \,\, \text{argmin}_{\bg^\mathfrak{T}_{\bx_{\small \circ\nxs},\textrm{\bf{n}}}} \norms{\bg^\mathfrak{T}_{\bx_{\small \circ\nxs},\textrm{\bf{n}}}}_{L^2}.}
\vspace{-1 mm}  
\eeq
The subscript $\mathfrak{T}$ indicates that the Tikhonov regularization is deployed to compute the LSM imaging functional.

\vspace{-1.5mm}
\subsubsection{The generalized linear sampling indicator}
In light of~\eqref{GLSMmg}, the GLSM-regularized solution $\bg^\mathfrak{G}_{\bx_{\small \circ\nxs},\textrm{\bf{n}}}$ to (\ref{Dff}) is obtained through solving the linear system
\vspace{-1 mm}
\beq \lb{min-DRJ} 
\Big( {\boldsymbol{\Lambda}}^{\! *}{\boldsymbol{\Lambda}}  + \gamma_{\bx_{\small \circ\nxs},\textrm{\bf{n}}} \exs  ({\boldsymbol{\Lambda}}^{\! *}{\boldsymbol{\Lambda}})^{\nxs\frac{1}{4}*} ({\boldsymbol{\Lambda}}^{\! *}{\boldsymbol{\Lambda}})^{\nxs\frac{1}{4}} + \delta \exs \gamma_{\bx_{\small \circ\nxs},\textrm{\bf{n}}}  \exs \boldsymbol{I}_{N_sN_\omega\times N_sN_\omega} \Big) \exs \bg^\mathfrak{G}_{\bx_{\small \circ\nxs},\textrm{\bf{n}}}  ~=~  {\boldsymbol{\Lambda}}^{\! *} \bPhi_{\bx_{\small \circ},\textrm{\bf{n}}},
\vspace{-1 mm}
\eeq
where $(\cdot)^*$ is the Hermitian operator, $\delta = 0.15\norms{\!{\boldsymbol{\Lambda}}\!}_{L^2}\!$ indicates the estimated magnitude of noise in data, and the regularization parameter 
\vspace{-1 mm}  
\beq\lb{Alph}
\gamma_{\bx_{\small \circ\nxs},\textrm{\bf{n}}} \,\,  = \,\, \frac{\eta_{\bx_{\small \circ\nxs},\textrm{\bf{n}}}}{\norms{\!{\boldsymbol{\Lambda}}\!}_{L^2} + \,\, \delta},
\vspace{-1 mm}
\eeq
wherein $\eta_{\bx_{\small \circ\nxs},\textrm{\bf{n}}}$ is as in~\eqref{lssm1}. As a result, $\bg^\mathfrak{G}_{\bx_{\small \circ\nxs},\textrm{\bf{n}}}$ is a $N_sN_\omega\times 1$ vector (or $N_sN_\omega\times M$ matrix for all the constructed right hand sides) identifying the distribution of wavefront densities over $S^{\text{inc}}$.  In this setting, the GLSM imaging functional is computed according to~(\ref{EIFn0}) as the following,
\vspace{-1 mm}
\beq\lb{GLSM-Dgs}
\mathfrak{G}(\bx_{\small \circ\nxs}) \,\, = \,\, \dfrac{1}{\sqrt{{\big(\bg^\mathfrak{G}_{\bx_{\small \circ\nxs}}, ({\boldsymbol{\Lambda}}^{\! *}{\boldsymbol{\Lambda}})^{\nxs\frac{1}{2}} \bg^\mathfrak{G}_{\bx_{\small \circ\nxs}}\big)} \exs+\,\, \delta \! \norms{\bg^\mathfrak{G}_{\bx_{\small \circ\nxs}}\!}^2_{L^2}}}, \qquad 
\textcolor{black}{
\bg^\mathfrak{G}_{\bx_{\small \circ\nxs}} \,\,\colon \!\!= \,\, \text{argmin}_{\bg^\mathfrak{G}_{\bx_{\small \circ\nxs},\textrm{\bf{n}}}} \norms{\bg^\mathfrak{G}_{\bx_{\small \circ\nxs},\textrm{\bf{n}}}}_{L^2}.} 
\vspace{-1 mm}
\eeq  

\vspace{-1.5mm}
\subsubsection{The linear sampling via direct inversion}
We observed that for $N_s = 8$, the operator $\boldsymbol{\Lambda}$ in (\ref{Dff}) is directly invertible owing to sparse sampling of $S^{\text{inc}}$. In this setting, one may also construct the LSM indicator from the directly inverted solution,
\vspace{-1 mm}  
\beq\lb{LSMD}
\mathfrak{L}(\bx_{\small \circ\nxs}) \,\, = \,\, \frac{1}{\norms{\bg_{\bx_{\small \circ\nxs}}\!}_{L^2}}, \qquad
\textcolor{black}{
\bg_{\bx_{\small \circ\nxs}} \,\,\colon \!\!= \,\, \text{argmin}_{\bg_{\bx_{\small \circ\nxs},\textrm{\bf{n}}}} \norms{\bg_{\bx_{\small \circ\nxs},\textrm{\bf{n}}}}_{L^2}, \qquad \bg_{\bx_{\small \circ\nxs},\textrm{\bf{n}}}~=~\boldsymbol{\Lambda}^{\!-1} \bPhi_{\bx_{\small \circ},\textrm{\bf{n}}}.}
\vspace{-1 mm}  
\eeq

A comparative study of the linear sampling indicators $\mathfrak{L}$ and $\mathfrak{L_T}$ is included in Section~\ref{RE}.    

The (generalized) linear sampling functionals canvas the support of 3PB-induced damage by achieving their highest values at sampling points that meet the support of newborn fractures $\Gamma$ (or micro-cracked process zones), while remaining near zero everywhere else within the sampling region $\mathcal{S} \backslash \Gamma$.    

\vspace{-1.5mm}
\subsubsection{The thresholded indicators}

On introducing 
\vspace{-1 mm}  
\[
\mathbbm{1}_{\mathfrak{I}}({\bx_{\small \circ}\nxs}) \,\, := \,\, 
\begin{cases}
1 \quad\quad\text{ if } \,\,\, \mathfrak{I}({\bx_{\small \circ}\nxs}) \,\,>\,\, \tau_{tol} \nxs\times\nxs \text{max}(\mathfrak{I})\\
0 \quad\quad\text{ otherwise}
\end{cases}\!\!\!\!\!\!, \qquad \mathfrak{I} \in \lbrace \mathfrak{L_T}, \mathfrak{L}, \mathfrak{G} \rbrace, \qquad \tau_{tol} \exs\in\,\, ]0\,\,\, 1[\, ,
\vspace{-1 mm}  
\]
the thresholded imaging functionals may be expressed as
\vspace{-1 mm}  
\begin{equation}\label{TiFM}
\tilde{\mathfrak{I}}({\bx_{\small \circ}\nxs}) \,\, := \,\, \mathbbm{1}_{\mathfrak{I}}({\bx_{\small \circ}\nxs}) \, \mathfrak{I}({\bx_{\small \circ}\nxs}), \qquad \mathfrak{I} \in \lbrace \mathfrak{L_T}, \mathfrak{L}, \mathfrak{G} \rbrace. 
\vspace{-1 mm}  
\end{equation}

\vspace{-1.5mm}
\section{Results and discussion}\lb{RE}

Following~\cite{pour2020}, the 3PB-induced damage is exposed by spraying acetone on the back of specimen in a neighborhood of the pre-manufactured notch. While evaporating, the acetone reveals the ``true" support of $\Gamma$ as illustrated in Fig.~\ref{GTC}. The latter is then compared with the reconstructed fractures $\Gamma_{\mathfrak{L}}$ and $\Gamma_{\mathfrak{G}}$ obtained by the LSM and GLSM indicators, respectively, according to Fig.~\ref{EIF}.         

\begin{figure}[!h]
\center\includegraphics[width=0.23\linewidth]{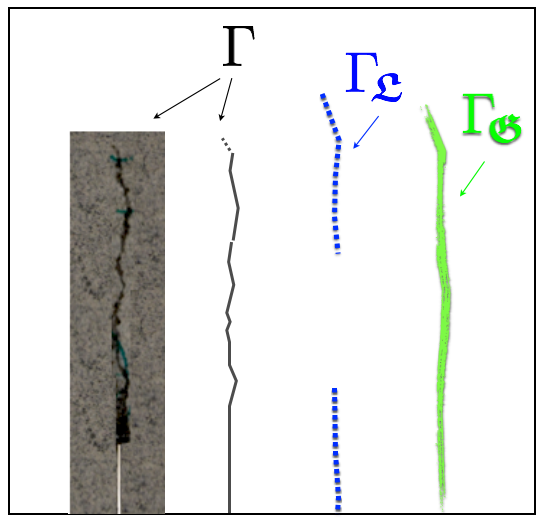} \vspace*{-3.5mm} 
\caption{Verification:~($\Gamma$)~3PB-induced fracture traced by acetone in a neighborhood of the pre-manufactured notch -- weak traces are indicated by the dashed line,~($\Gamma_{\mathfrak{L}}$)~recovered support of $\Gamma$ by way of the classical linear sampling indicator $\mathfrak{L}$, and~($\Gamma_{\mathfrak{G}}$)~reconstructed damage via the generalized linear sampling indicator $\mathfrak{G}$. $\Gamma_{\mathfrak{L}}$ and $\Gamma_{\mathfrak{G}}$ are extracted from Fig.~\ref{EIF}.}\lb{GTC}
\vspace*{-2mm}
\end{figure}

\vspace*{-1.5mm}
\subsection{Full aperture reconstruction}

The spectrum of scattered displacement data $F(\bv)(\bxi_i,\omega_\ell)$ measured at 144 observation points $\bxi_i \in S\obs \!=\! \bigcup_{\iota=1}^4\mathcal{G}_\iota$, $i = 0, \ldots, 143$, for ten frequencies $\omega_\ell = 27, 28, \ldots, 36$kHz, and eight source locations on $S^{\text{inc}} = \lbrace {s}_1, s_2, \ldots, s_8 \rbrace$ are deployed to compute the (G)LSM imaging functionals $\mathfrak{L_T}$, $\mathfrak{L}$, and~$\mathfrak{G}$ according to~\eqref{LSM2},~\eqref{LSMD}, and~\eqref{GLSM-Dgs}, respectively. Recall that the sampling region is a $29$cm $\nxs\!\times\!\nxs$ $29$cm square in the middle of specimen. The resulting distributions are shown in Fig.~\ref{EIF}. As mentioned earlier, the (G)LSM imaging functionals assume their highest values in the vicinity of hidden scatterers $\Gamma$. It is worth mentioning that the caustics featured in the reconstructed maps of Fig.~\ref{EIF} are mostly governed by~(i)~illuminating wavelength,~(ii)~geometric symmetries of the domain,~(iii)~arrangement of sources and receivers, and~(iv)~mathematical properties of the associated cost functionals. Their intensity typically decreases when the source and measurement aperture along with the number of sources and receivers increase. An in-depth analysis of such focal regions for a related indicator known as the topological sensitivity is provided in~\cite{Fatemeh2015(2)}.

Fig.~\ref{EIF} also includes the $60$\% thresholded maps $\tilde{\mathfrak{L}}_{\mathfrak{T}}$, $\tilde{\mathfrak{L}}$, and~$\tilde{\mathfrak{G}}$ furnishing the support of sampling points $\bx_{\circ\nxs}$ that satisfy $\mathfrak{I}({\bx_{\small \circ}\nxs}) > 0.6 \nxs\times\nxs \text{max}(\mathfrak{I})$, $\mathfrak{I} \in \lbrace \mathfrak{L_T}, \mathfrak{L}, \mathfrak{G} \rbrace$, according to~\eqref{TiFM}. These results are used to approximate the support of damage $\Gamma_{\mathfrak{L}}$ and $\Gamma_{\mathfrak{G}}$ by the midline through the thresholded damage zone as shown in the figure. It is instructive to compare $\Gamma_{\mathfrak{L}}$ and $\Gamma_{\mathfrak{G}}$ with the ``true" fracture boundary $\Gamma$ from Fig.~\ref{GTC} -- also included as an inset in Fig.~\ref{EIF}. Observe that both LSM and GLSM reconstructions indicate that the damage zone has advanced slightly further in the specimen compared to $\Gamma$. This may be justified by noting that acetone -- used to recover $\Gamma$, detects only the sufficiently penetrable interfaces which may not include the tight contacts in the near tip region. 

A comparative analysis of Fig.~\ref{EIF} indicates that the LSM functionals $\mathfrak{L_T}$ and $\mathfrak{L}$ result in quite similar reconstructions. In light of $\tilde{\mathfrak{L}}_{\mathfrak{T}}$ and $\tilde{\mathfrak{L}}$, however, observe that when the scattering operator $\boldsymbol{\Lambda}$ is invertible -- here, thanks to the sparse sampling of $S^{\text{inc}}$, the direct-inversion-based operator $\mathfrak{L}$ leads to a ``cleaner" reconstruction. In other words, the Tikhonov regularization, owing to its approximate nature, may intensify the caustics giving rise to a ``noisy" $\mathfrak{L_T}$ reconstruction. Henceforth, we focus on the LSM maps constructed via direct inversion. The GLSM indicator $\mathfrak{G}$, on the other hand, successfully recovers the entire damage zone with a sharp localization in a neighborhood of $\Gamma$ and remarkably diminished reconstruction artifacts. This may be attributed to:~(a)~rigorous nature of the GLSM imaging functional which does not involve approximations underlying the LSM indicator, and~(b)~strong convexity of the GLSM cost functional~\eqref{GLSMmg}, see e.g.,~\cite[Theorem 4.3]{pour2019}.         

\begin{figure}[!h]
\center\includegraphics[width=0.65\linewidth]{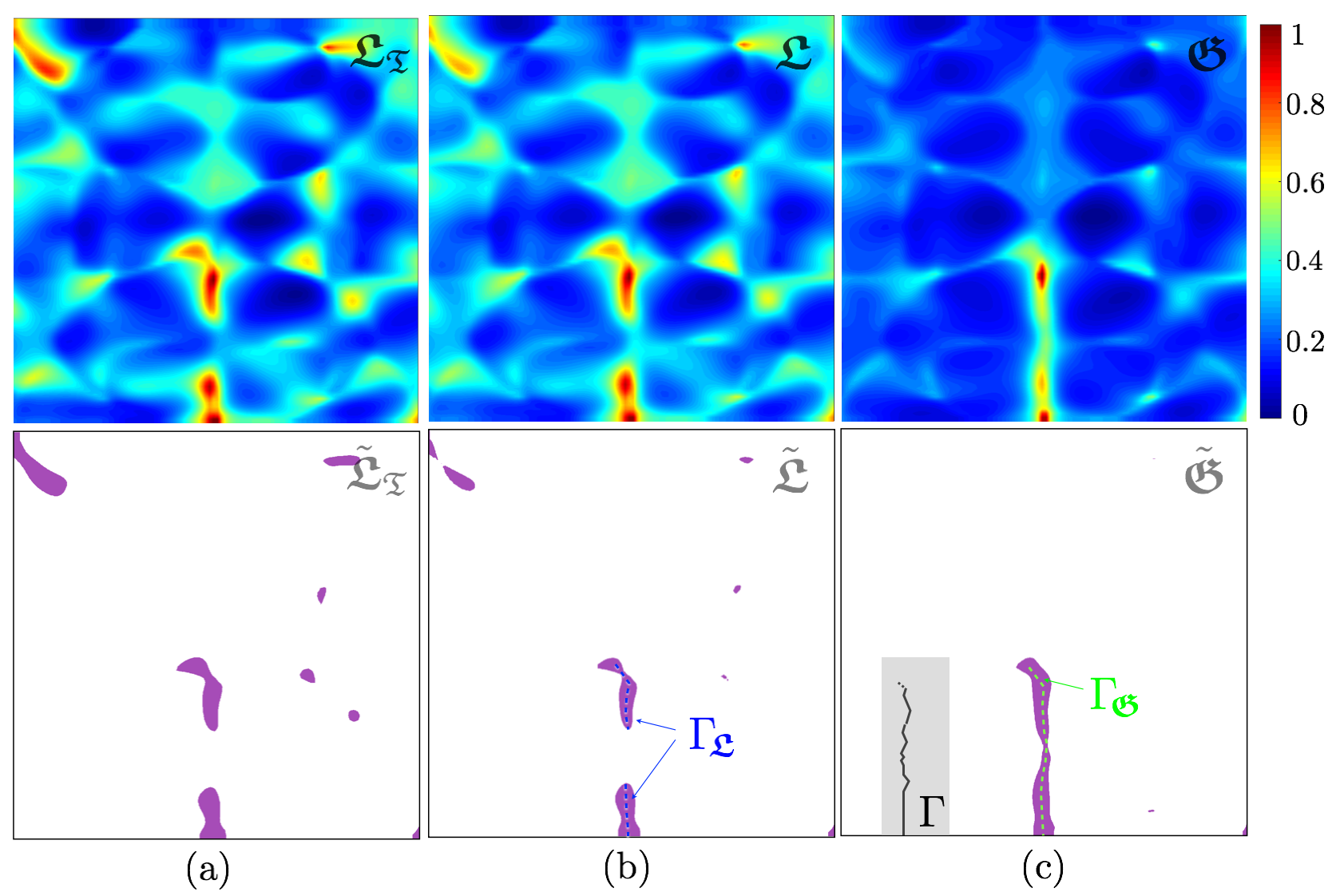} \vspace*{-2.5mm} 
\caption{(Generalized) linear sampling indicators:~(a)~LSM indicator $\mathfrak{L_T}$~\eqref{LSM2} computed via the Tikhonov regularization in the sampling region -- a $29$cm $\nxs\!\times\!\nxs$ $29$cm square in the middle of specimen, and the associated thresholded indicator $\tilde{\mathfrak{L}}_{\mathfrak{T}}$~\eqref{TiFM} with $\tau_{tol} = 0.6$,~(b)~LSM map $\mathfrak{L}$~\eqref{LSMD} obtained via direct inversion and the corresponding $\tilde{\mathfrak{L}}$ thresholded at $60\%$, and~(c)~GLSM indicator map $\mathfrak{G}$~\eqref{GLSM-Dgs} and the affiliated $\tilde{\mathfrak{G}}$ similarly truncated at $60\%$. The inset shows the ``ground-truth" support of $\Gamma$ from Fig~\ref{GTC}. Here, full ultrasonic data is deployed for the reconstruction according to Fig.~\ref{Exp-sch}(b) where $S^{\text{inc}} = \lbrace {s}_1, s_2, \ldots, s_8 \rbrace$ and $S\obs = \bigcup_{i=1}^4\mathcal{G}_i$ involving 144 measurement points.} \lb{EIF}
\vspace*{-2.5mm}
\end{figure} 

\begin{figure}[!h]
\center\includegraphics[width=0.875\linewidth]{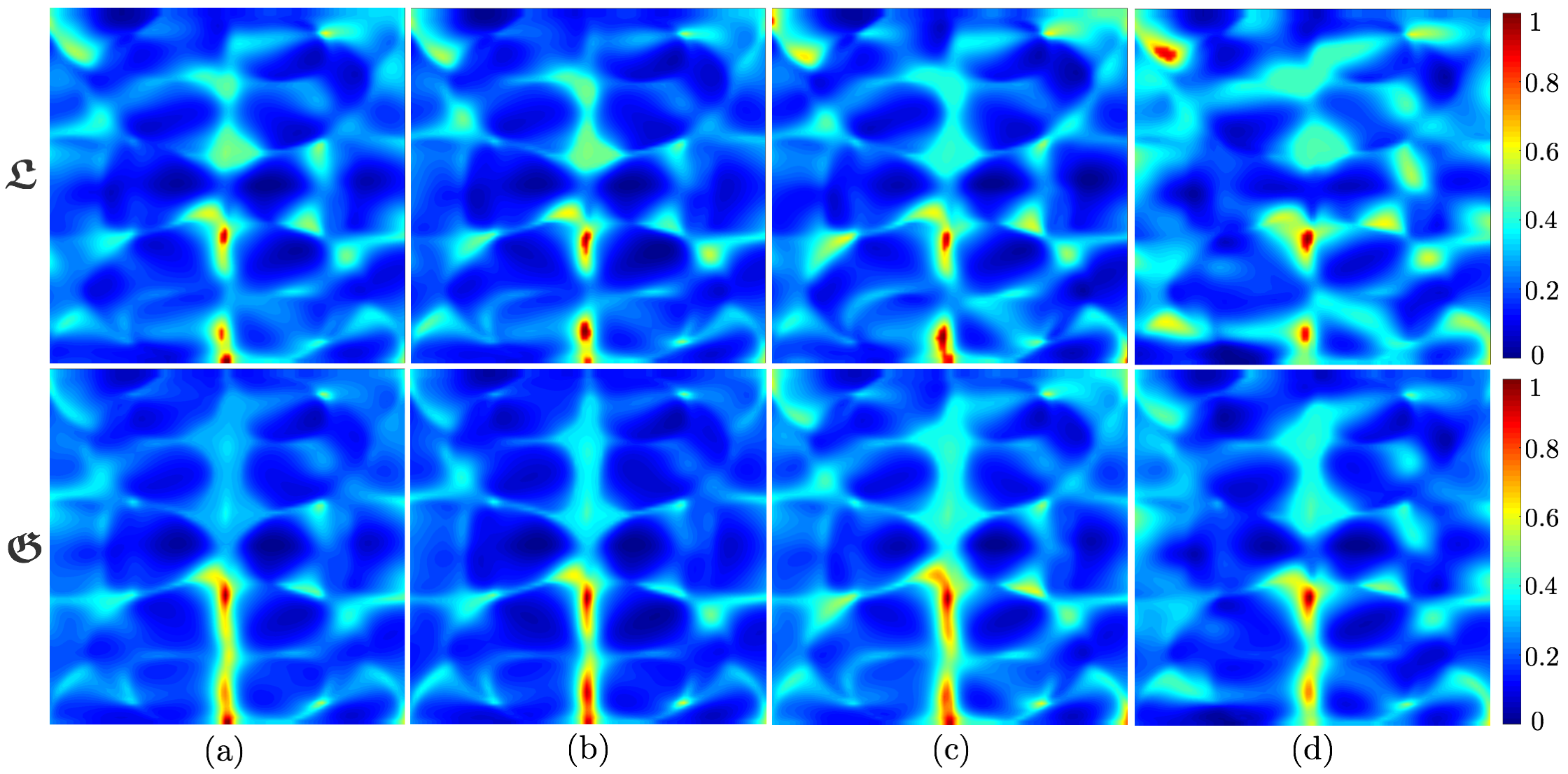} \vspace*{-2.5mm} 
\caption{LSM (top) versus GLSM (bottom) indicator maps computed from reduced data where $S^{\text{inc}} = \lbrace {s}_1, s_2, \ldots, s_8 \rbrace$, while $S\obs = \bigcup_{i=1}^4\mathcal{G}_i$ is uniformly downsampled by a factor of:~(a)~three (corresponding to $N_p = 48$ measurement points),~(b)~five ($N_p = 28$),~(c)~seven ($N_p = 20$), and~(d)~nine ($N_p = 16$).} \lb{REIF}
\vspace*{-2.5mm}
\end{figure}

\vspace*{-1.5mm}
\subsection{Reconstruction from reduced data}

To examine the performance of (G)LSM indicators with sparse data, the measurement points on $S\obs$ are uniformly downsampled by a factor of $\beta \in \lbrace 3, 5, 7, 9 \rbrace$, so that a respective set of $N_p \in \lbrace 48, 28, 20, 16 \rbrace$ data points are used for the reconstruction -- compared to $N_p = 144$ in Fig.~\ref{EIF}. The resulting $\mathfrak{L}$ and $\mathfrak{G}$ distributions are shown in~Fig.~\ref{REIF} for all $\beta$. Observe that while the GLSM indicator remains robust against downsampling, owing to its rigorous nature, the LSM indicator fails to retrieve the damage zone from sparse data, especially when $N_p \leqslant 20$. This is more evident in the 60\% thresholded maps $\tilde{\mathfrak{L}}$ and $\tilde{\mathfrak{G}}$ shown in~Fig.~\ref{TEIF}. Note that as the number of data points $N_p$ decreases,~(a)~caustics and reconstruction artifacts intensify in both maps which is rather expected in light of~\cite{Fatemeh2015(2)}, and~(b)~image resolution decreases in the GLSM maps.   

\vspace*{-1.5mm}
\subsubsection*{Partial source and ``viewing" aperture}

It is common in practice that a specimen is inaccessible from one side or, to the contrary, is only accessible from one side for ultrasonic testing. Imaging in such configurations are investigated in Fig.~\ref{OEIF}. In the top row, the specimen is assumed inaccessible from below for both excitation and measurement, and thus, the reconstruction is performed using data on three sides of the boundary $S\obs = \bigcup_{i=1}^3\mathcal{G}_i$ involving 99 measurement points for six source locations -- i.e., $S^{\text{inc}} = \lbrace {s}_1, s_2, s_3, s_4, s_5, s_8 \rbrace$. The LSM and GLSM indicators are able to recover most of the damage support. However, the GLSM functional appear to be more robust with less pronounced artifacts. In the bottom row, the specimen is presumed to be merely accessible from the top for ultrasonic illumination and sensing. In this setting, $\mathfrak{L}$ and $\mathfrak{G}$ are computed using limited data involving four ultrasonic sources on top $S^{\text{inc}} = \lbrace {s}_1, s_2, s_3, s_4 \rbrace$, and 45 measurement points on $S\obs = \mathcal{G}_2$. In this case, the GLSM map successfully recovers the damage zone, while the LSM distribution canvases only a subset of the fracture support.

\begin{figure}[!h]
\center\includegraphics[width=0.85\linewidth]{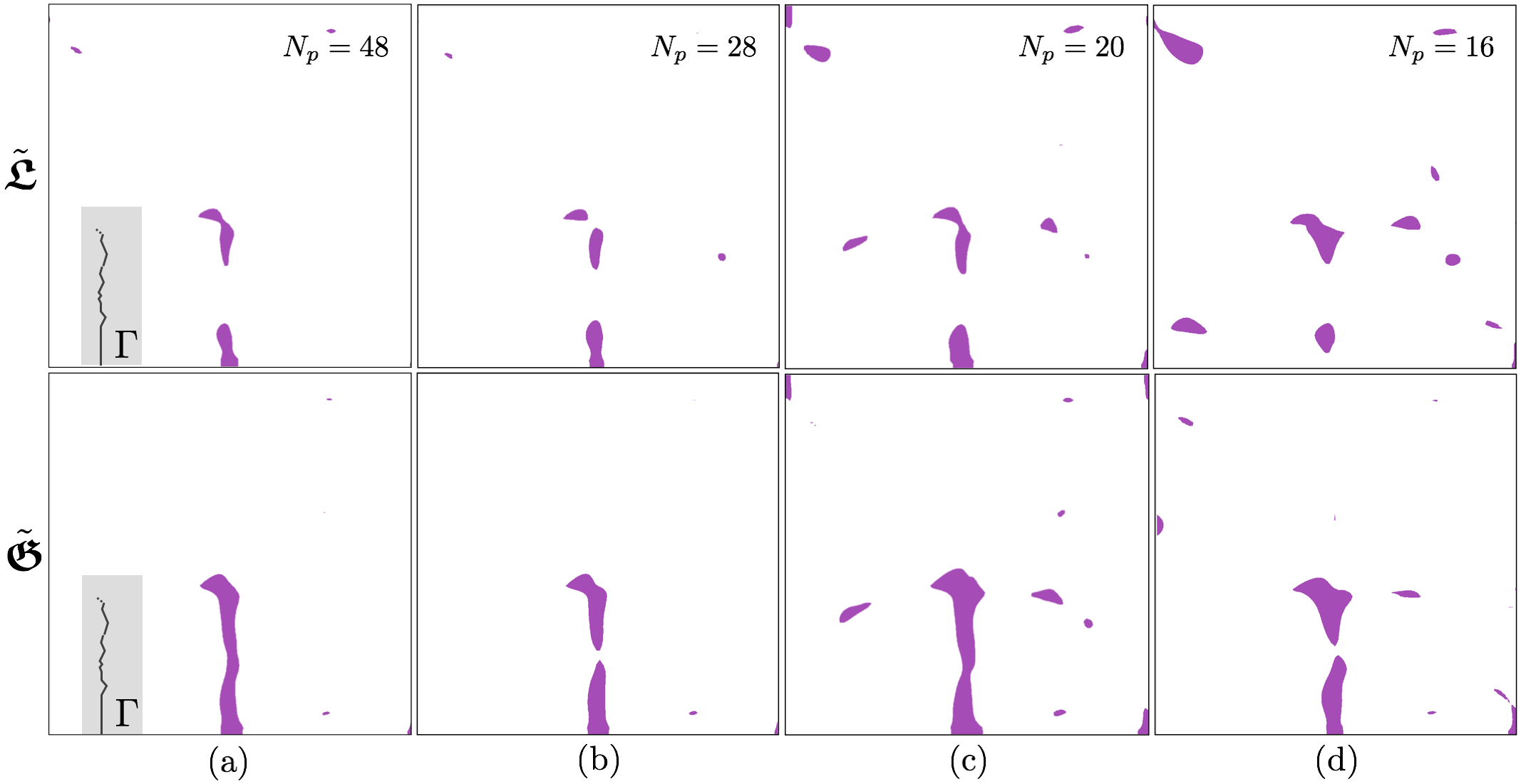} \vspace*{-2.5mm} 
\caption{Thresholded indicator maps $\tilde{\mathfrak{L}}$ (top) and $\tilde{\mathfrak{G}}$ (bottom) associated with the LSM and GLSM distributions of Fig.~\ref{REIF}. The number of (downsampled) measurement points $N_p$ is specified for every column~(a)-(d). The insets in~(a)~are from Fig~\ref{GTC}, providing the ``ground-truth" for the 3PB-induced fracture $\Gamma$. With reference to~\eqref{TiFM}, the threshold in all cases is $\tau_{tol} = 0.6$.} \lb{TEIF}
\vspace*{-2.5mm}
\end{figure}

\begin{figure}[!h]
\center\includegraphics[width=0.7\linewidth]{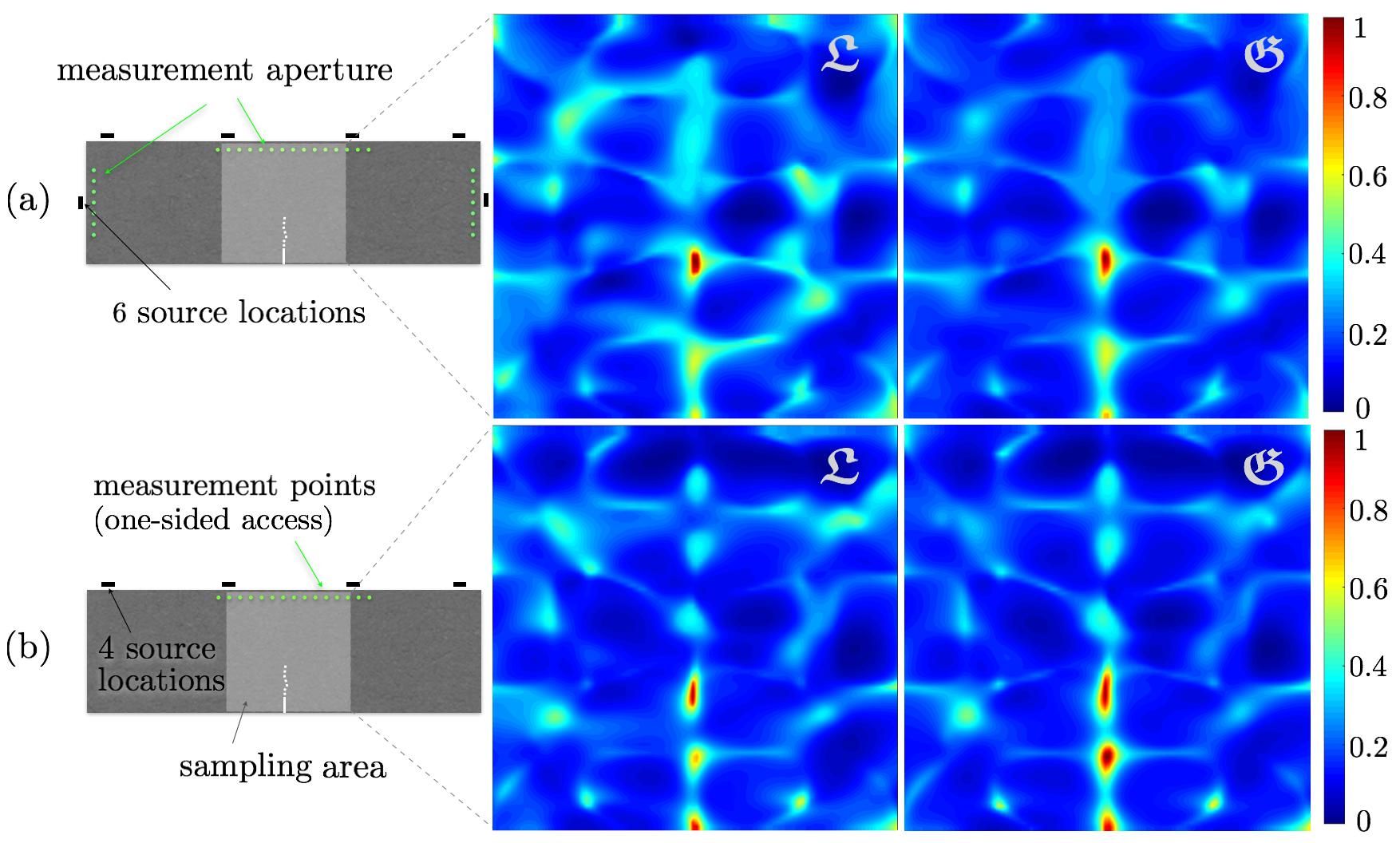} \vspace*{-2.5mm} 
\caption{Partial-aperture tomography: LSM ${\mathfrak{L}}$ (middle) and GLSM ${\mathfrak{G}}$ (right) indicator maps computed using limited data involving: (a) six ultrasonic sources on $S^{\text{inc}} = \lbrace {s}_1, s_2, s_3, s_4, s_5, s_8 \rbrace$ and 99 measurement points on $S\obs = \bigcup_{i=1}^3\mathcal{G}_i$ as shown in the top left panel, and (b) four sources on $S^{\text{inc}} = \lbrace {s}_1, s_2, s_3, s_4 \rbrace$, and 45 points on $S\obs = \mathcal{G}_2$ as depicted in the bottom left panel.} \lb{OEIF}
\vspace*{-2.5mm}
\end{figure}

\section{Conclusions}
 
\noindent An experimental and data analysis framework is developed for in-situ waveform tomography of damage in elastic components. To this end, we take advantage of the recently established generalized linear sampling indicator for non-iterative, full-waveform reconstruction of a mode I fracture, induced via three-point bending, in a granite specimen using boundary observations of scattered ultrasonic waveforms. In this vein, transient waves ranging from 20 to 40kHz are induced in the sample, and thus generated velocity responses are monitored by a 3D scanning laser Doppler vibrometer over the domain's external boundary, which upon suitable signal processing furnish the spectra of scattered displacement fields over the designated scanning grid. Such sensory data are then deployed to compute the GLSM maps along with the classical LSM indicators for a comparative analysis. The results are verified against in-situ observations and shown to be successful in recovering the damage support. The GLSM, however, leads to a sharper localization and remarkably cleaner maps -- with less-pronounced reconstruction artifacts. It is further demonstrated that the GLSM remains robust with reduced i.e., spatially downsampled data, as well as partial-aperture data e.g., when access to specimen for excitation and sensing is limited.  In this study, the data inversion procedure is adapted for a multifrequency reconstruction. Given the transient nature of data, it would be interesting to extend the theory for a direct implementation of this approach in the time domain. In this setting, a broadband dataset opens the door toward an in-depth analysis of multi-scale fracture networks in a damage zone.

\section*{Acknowledgments} 

\noindent The experimental campaign was conducted in the Department of Civil, Environmental \& Geo- Engineering at the University of Minnesota. The author kindly acknowledges the comprehensive support provided by Professor Bojan Guzina in the course of experiments. Special thanks are due to Roman Tokmashev for his assistance with the experiments. This study was funded by the University of Colorado Boulder through FP's startup. This work utilized resources from the University of Colorado Boulder Research Computing Group, which is supported by the National Science Foundation (awards ACI-1532235 and ACI-1532236), the University of Colorado Boulder, and Colorado State University.   

\bibliography{inverse,crackbib}

\end{document}